\documentclass[twocolumn]{svjour3}      
\smartqed  

\usepackage[utf8]{inputenc} 
\usepackage{hyperref}
\usepackage{subfig}
\usepackage{amsmath}
\usepackage{booktabs}
\usepackage{graphicx}
\usepackage{marginnote}
\usepackage{bm}
\usepackage{caption}
\usepackage[ruled,vlined,onelanguage]{algorithm2e}
\usepackage[noend]{algpseudocode}

\usepackage[dvipsnames]{xcolor}

\usepackage{floatrow}
\usepackage{graphbox}

\begin{document}


\title{On the Modeling and Simulation of Portfolio Allocation Schemes: an Approach based on Network Community Detection}

\author{Stefano Ferretti}
\institute{S. Ferretti \at
              Department of Pure and Applied Sciences,\\ University of Urbino ``Carlo Bo'',\\ Urbino, Italy\\
              \email{stefano.ferretti@uniurb.it}
}
\maketitle


\begin{abstract} 
We present a study on portfolio investments in financial applications. We describe a general modeling and simulation framework and study the impact on the use of different metrics to measure the correlation among assets. 
In particular, besides the traditional Pearson's correlation, we employ the Detrended Cross-Correlation Analysis (DCCA) and Detrended Partial Cross-Correlation Analysis (DPCCA). 
Moreover, a novel portfolio allocation scheme is introduced that treats assets as a complex network and uses modularity to detect communities of correlated assets. Weights of the allocation are then distributed among different communities for the sake of diversification. Simulations compare this novel scheme against Critical Line Algorithm (CLA), Inverse Variance Portfolio (IVP), the Hierarchical Risk Parity (HRP). Synthetic times series are generated using the Gaussian model, Geometric Brownian motion, GARCH, ARFIMA and modified ARFIMA models.
Results show that the proposed scheme outperforms state of the art approaches in many scenarios.
We also validate simulation results via backtesting, whose results confirm the viability of the proposal.
\end{abstract}



\section{Introduction}

Recent advances in financial technologies (fintech) and decentralized finance (DeFi) are revolutionizing the way we invest and manage our wealth. For instance, the advent of cryptocurrencies, ICOs and related DeFi contexts, has surely reshaped the patterns of investments, as well as the audience of people interested in investing their money.
This new and vibrant scenario fosters novel opportunities and offers elements that suggest revising and improving the techniques to allocate investments on different possible assets and stocks (portfolio management), based on data science and data analysis techniques. 
This involves three main aspects related to i) the way data traces are analyzed, ii) how the portfolio allocation is performed, iii) which techniques are used to assess whether the portfolio allocation schemes can effectively perform in various scenarios.

As concerns data trace analysis, in the last few years novel metrics have been introduced, that try to better characterize the correlations among different data series, which cannot be considered as produced by stationary processes. These approaches are usually based on detrended fluctuation analyses. In particular, the Detrended Cross Correlation Analysis (DCCA) and Detrended Partial Cross Correlation Analysis (DPCCA) seem to be promising metrics 
\cite{Oh2011StatisticalPO,PhysRevE.83.046121,GUEDES201738,RePEc:arx:papers:0709.0281,NIPS2017_ffeabd22}.

Portfolio allocation is a main research topic, that has been studied in depth in the last decades, especially by the economic and financial community. However, novel studies showed that the contribution of techniques based data and computer science can be quite beneficial. In fact, they promote the design of innovative allocations schemes based on machine learning and data analysis \cite{deprado,WU2021668,MEHLAWAT2021348,KIM2021103468,ZHANG2020548,KWAK2021115298}. For instance, \cite{deprado} presents a portfolio asset allocation scheme that exploits clustering techniques. Moreover, \cite{KWAK2021115298,ANDERSSON2021126399} are examples of papers where a deep learning framework is proposed to optimize some portfolios management aspects. Needless to say, due to the novelty of these techniques, there is room for improvement.

In order to foster the design of novel portfolio allocation schemes, there is the need for a framework that allows to effectively study how these devised schemes perform in different scenarios.
Indeed, one approach to studying different portfolio allocation strategies is based on backtesting, i.e., taking historical data traces of different asset and stock prices and trying to simulate how the scheme performs over these traces. 
The principal danger of this common approach is that it can generate statistical overfitting \cite{CESARI2003987}. The computational capabilities of modern computers enable the analysis of thousands of variations of a given strategy, thus allowing to perfectly tune the hyper-parameters of the schemes over such traces. 
To avoid overfitting, a viable solution consists in performing (Monte-Carlo) simulation analyses. In essence, multiple pseudo-random data trace generation is exploited to study the devised portfolio allocation schemes. Studying the behavior of a scheme over hundreds of (randomly generated) simulation scenarios allows obtaining a wide and general idea about the performance of the scheme in that type of scenario.
Backtesting can be finally used as a final test set, to confirm the viability of the devised approach.

In this paper, we try to give a contribution in these three dimensions. In particular, first, we present a Monte Carlo simulation framework, in which different portfolio allocation strategies are compared. Different typologies of data traces are studied to widely assess the schemes. 
Then, as a further test, we provide results of a more traditional backtest approach. 
Backtest outcomes confirm the results obtained from simulations. 

Second, we implement different variants of the considered schemes, that are based on different metrics used to assess the correlation among data traces, i.e., the traditional Pearson correlation coefficient, DCCA and DPCCA. 
In particular, we plug these metrics into the allocation schemes, to analyze whether different behaviors are obtained and if some among these metrics perform better than others.

Third, we propose a novel portfolio allocation scheme that resorts to the idea that a group of assets can be treated as a complex network. 
Thus, we represent the set of assets as a graph and use network modularity to find communities of assets, based on their level of correlation. Weights of the allocation are equally distributed among these communities. While the approach is very simple and naive, results show that in most cases this approach outperforms other well-known strategies. This confirms that the use of complex network theory, and data science techniques in general, are important tools to consider in portfolio analysis applications.

The remainder of this paper is organized as follows.
Section \ref{sec:back} provides a general background, needed to easily follow the discussion provided in the rest of the paper. Section \ref{sec:schemes} introduces the main terminology and notation, as well as the different portfolio allocation schemes that are evaluated in the subsequent sections.  
Section \ref{sec:eval} describes the simulation framework and the evaluation study based on different types of simulation employed to generate the different corpuses of data traces. 
In this section, we also introduce the metrics employed to evaluate the compared schemes. Section \ref{sec:res} presents and discusses the results obtained using simulation, while Section \ref{sec:backtest} presents results from a backtest analysis. 
Section \ref{sec:disc} provides a discussion on the results coming from the simulations and backtest, to provide a final overview on the performance of the considered approaches in different scenarios.
Finally, Section \ref{sec:conc} provides some concluding remarks.

\section{Background}
\label{sec:back}

In this section, a general background is introduced. Clearly enough, we cannot cover all the aspects related to modern portfolio theory. The interested reader can find a variety of interesting resources on the topic, e.g., \cite{markowitz,Elton_2014a}.

\subsection{Modern Portfolio Theory}
In 1952, Markowitz published a seminal work introducing an investment theory based on mean-variance optimization \cite{markowitz}. 
Let's assume that you have a certain amount of money to invest; a given set of possible assets, the problem is to find the optimal asset allocation, i.e., how much you are going to invest in each asset. 
Portfolio allocation schemes cope with this issue, trying to identify the best trade-off between the expected return and risk (measured as the variance of returns).
Indeed, some assets might have a high volatility that can result in higher profits, as well as in higher associated risks. 

Since the goal is to distribute the allocation of investments into variegated assets, these approaches employ a forecast of the covariance matrix of the expected returns, in order to understand if different assets are correlated or not.
Notable risk-based portfolio allocation methods that rely on such covariance forecasts are the minimum variance \cite{Clarke10}, maximum diversification \cite{Choueifaty40}, equal risk budget \cite{Leote12},
equal risk contribution\cite{Maillard10}. 
Usually, the covariance of the expected returns are estimated using the covariance of a sample of previous returns.

In Section \ref{sec:schemes}, we will introduce the main notation and go deeper into some details on the specific schemes which are considered in this study.

\subsection{Detrended Fluctuation Analysis}\label{sec:dpcca}
A main problem here, in the end, is to properly analyze times
series related to assets returns. 
The difficulty is due to the fact that a financial system is a complex system, influenced by a multitude of concurrent and usually unknown factors.
Not only, as it happens in many types of other complex systems, e.g., climatology, biology, etc., when we look at many time series, we find that their fluctuations may exhibit cross-correlation characteristics.
As a demonstration, in \cite{Oh2011StatisticalPO}, Stanley et al.~revealed the existence of cross-correlation properties among stocks in the Korean market. Moreover, in \cite{PhysRevE.83.046121}, 48 financial indices were considered and long-range power-law cross-correlations in their returns have been identified.

All this reveals a limitation on the typical use of the traditional Pearson's correlation coefficient on financial data traces, whose use is justified to represent a linear correlation between two time series, which are both assumed to be stationary.  
To address the drawbacks of Pearson's correlation, the Detrended Cross-Correlation Analysis (DCCA) \cite{GUEDES201738,RePEc:arx:papers:0709.0281} and the Detrended Partial-Cross-Correlation Analysis (DPCCA) have been recently introduced \cite{NIPS2017_ffeabd22}. 
These approaches are a generalization of the method of Detrended Fluctuation Analysis (DFA) for non-stationary time series \cite{PhysRevE.49.1685}.

\subsubsection{Detrended Cross-Correlation Analysis (DCCA)}\label{sec:dcca}
The Detrended Cross-Correlation coefficient $\rho_{\textit{DCCA}}(n)$ is a measure aimed at  quantifying the level of cross-correlation between non-stationary time series $\{x_i\}, \{y_i\}$. It is defined as the ratio between the detrended covariance function $F^2_{xy}$ and the detrended variance functions $F_{xx}(n)$, $F_{yy}(n)$ of the two series, i.e.
\begin{equation}\label{eq:rho}
\rho_{\textit{DCCA},xy}(n)=\frac{F^2_{xv}}{F_{xx}(n) F_{yy}(n)}.
\end{equation}
The value of $\rho_{\textit{DCCA}}(n)$\footnote{With some abuse of notation, when possible we omit to specify the names of the time series, thus preferring $\rho_{\textit{DCCA}}(n)$ to $\rho_{\textit{DCCA},xy}(n)$, for the sake of simplicity.} ranges between $-1$ and $1$. A value of $\rho_{\textit{DCCA}}(n)=0$ means there is no cross-correlation, while values $-1$ or $1$ reveal a perfect negative or positive correlation.
Such measure can be obtained through the following algorithm \cite{GUEDES2019121286}: 

\paragraph{Step I}
Given the two times series $\{x_i\}, \{y_i\}, i=1,\ldots,T$, we create two integrated series
$$X_k=\sum_{i=1}^k x_i - \langle x \rangle,\ Y_k=\sum_{i=1}^k y_i - \langle y \rangle,$$
with $\langle x \rangle, \langle y \rangle$ being the mean value of each time series, and $k=1,\ldots,T$.

\paragraph{Step II}
Divide $\{X_k\}, \{Y_k\}$ into  overlapping boxes of equal length $n$, being $n$ a parameter to set.

\paragraph{Step III}
In each box $j$, a linear fit is performed (by least-squares fit) for each series, here denoted $\{\tilde{X}_{k,j}\}, \{\tilde{Y}_{k,j}\}$.
The covariance of the residuals in each box (length n) is 
$$f^2_{xy}(n,j)=\frac{1}{(n+1)} \sum_{k=j}^{j+n} (X_k - \tilde{X}_{k,j})(Y_k - \tilde{Y}_{k,j}).$$

\paragraph{Step IV}
The detrended covariance function $F^2_{xy}(n)$ is calculated as the mean of the covariance $f^2_{xy}$ over all the $(T-n)$ boxes
$$F^2_{xy}(n)=\frac{1}{(T-n)} \sum_{i=1}^{T-n} f^2_{xy}(n,i).$$

\paragraph{Step V}
Having the values of $F^2_{xy}(n)$, and being $F_{xx}(n)=\sqrt{F^2_{xx}(n)}$, we can compute $\rho_{\textit{DCCA},xy}(n)$ through Equation (\ref{eq:rho}).

Clearly enough, if we have a set of $N$ time series (representing asset returns, for instance), we can create a matrix of DCCA coefficients, i.e., 
$$\bm{\rho}_{\textit{DCCA}}(n) = [\rho_{\textit{DCCA},ij}(n)],$$ 
with $i,j=1,\ldots,N$.

As already mentioned, this metrics has been already studied in a wide range of different application scenarios where data series are used \cite{PhysRevE.83.046121,RePEc:arx:papers:0709.0281,NIPS2017_ffeabd22,RePEc:eee:phsmap:v:392:y:2013:i:8:p:1756-1761,RePEc:eee:phsmap:v:402:y:2014:i:c:p:291-298,RePEc:eee:phsmap:v:390:y:2011:i:4:p:614-618} and even in economic contexts \cite{GUEDES201738,GUEDES2019121286,FERREIRA2020123803}. Thus, it becomes interesting to understand if this is a viable metrics to use in portfolio asset allocation.

\subsubsection{Detrended Partial Cross-Correlation Analysis (DPCCA)}
Detrended Partial Cross-Correlation Analysis (DPCCA) is an extension of DCCA. Its aim is to combine the advantages of DCCA and partial correlation to further improve the ability to quantify the relation between non-stationary data series. That is, similarly to DCCA, DPCCA should remove the effects of non-stationarity and provide information on the cross-correlation. Moreover, DPCCA should allow investigating the correlations of multiple series in a complex system and find their intrinsic relations \cite{Yuan2015}.

To measure it, given the matrix of DCCA coefficients $\bm{\rho}_{\textit{DCCA}}(n)$, we must invert it, obtaining $\bm{C}(n)=\bm{\rho}^{-1}_{\textit{DCCA}}(n)$. Then, the coefficients of the DPCCA are measured as
$$\rho_{\textit{DPCCA},xy}(n)=\frac{-C_{xy}(n)}{\sqrt{C_{xx}(n) C_{yy}(n)}}$$
being $C_{xy}(n)$ the $(x,y)$-th element of the matrix $\bm{C}(n)$.

In the rest of this work, we will exploit these two correlation metrics as an alternative to the classic Pearson's correlation, within the portfolio allocations schemes. An implementation of DCCA and and DPCCA is available in \cite{NIPS2017_ffeabd22,git-dpcca}

\section{The compared schemes}\label{sec:schemes}

In this section, we overview the portfolio allocation schemes that we use in our study. 
Three of them represent the state of the art of modern portfolio theory, i.e., Critical Line Algorithm (CLA), Inverse Variance Portfolio (IVP) and Hierarchical Risk Parity (HRP). 
Furthermore, we propose a novel scheme that exploits complex network theory and network modularity as the main elements to categorize and cluster different assets. 

All the approaches exploit the standard statistical covariance and correlation as the means to identify those assets that are similar.
While this represents a viable and reasonable approach, it forces the analyst to make the (strong) assumption that asset trends are stationary. Moreover, these metrics are tied to the returns time series, not taking into consideration other possible external information that can be easily inferred today, thanks for instance to novel data science and machine learning techniques. Examples go from a simple characterization of assets based on their typology (e.g., hi-tech companies, banking, healthcare, etc.), to the use of sentiment analysis techniques based, for instance, on the analysis of social networks \cite{DeMichele2019,ferFurMont}. These considerations suggest that there is room for improvement, in this sense.

Trying to follow this idea, we will consider some variants of the classic approaches. In particular, we plug into the schemes three different correlation metrics, i.e., Pearson's correlation, DCCA and DPCCA, to assess their performance. 

\subsection{Main Terminology and Notation}

We consider a generic portfolio composed of $N$ risky assets $\mathbf{a}=(a_1, \ldots, a_N)$. The return of an asset $a_i$, at time $t$ is denoted as $r_{t,i}$\footnote{When time is not important, for the sake of legibility we will omit the $t$ subscript.}. Weights represent the share of wealth invested in assets. 
We denote with $\mathbf{w} = (w_1, \ldots, w_N)$ the vector of weights associated with the set of assets $\mathbf{a}$.
Thus, the expected return on the portfolio is $\mathbf{w^T r}$.

The $N \times N$ covariance matrix of the returns $\mathbf{r} = (r_1, \ldots ,r_N)$ is denoted by $\mathbf{\Sigma}$.
Given the assets' time series, the covariance matrix is obtained by taking a window of the assets’ return time series and by performing the calculation of the covariance in those time intervals.
The standard deviation of returns is reported in the main diagonal of the covariance matrix, i.e.,
$\boldsymbol\sigma = \sqrt{\text{diag}(\mathbf{\Sigma})}$. Pearson's correlation matrix can be measured as $\boldsymbol\sigma^{-1} \mathbf{\Sigma} \boldsymbol\sigma^{-1}$. Conversely, the DCCA and DPCCA correlation measures are calculated as described in the previous section.
Since we are contrasting different measures of correlation, we will refer to Pearson's correlation, DCCA or DPCCA coefficients, depending on the specific variant of the scheme under investigation. 
In the experimental evaluation, these variants are referred as ``cov", ``dcca" and ``dpcca", respectively.

\subsection{Critical Line Algorithm (CLA)}
The Critical Line Algorithm (CLA) is the one introduced by Markowitz in \cite{markowitz}. It
solves a quadratic optimization problem with  constraints on each weight \cite{a6010169}. 
The approach focuses on deriving an efficient portfolio that yields the maximum return for a minimum risk (or volatility). 
More specifically, the approach tries to solve the following optimization
\begin{flalign*}
 & &  \text{minimize }   & \mathbf{w}^T \boldsymbol\Sigma \mathbf{w}   &\\
 & &  \text{subject to } & l_i \leq w_i \leq u_i          &\\
 & &                    & \sum_{i=1}^N w_i = 1           &\\
 & &                    & \sum_{i=1}^N w_i r_i = r_p &
\end{flalign*}
where $l_i, u_i$ are lower and upper bounds for the weights $w_i$, a constraint is present to ensure that the sum of the weights is equal to $1$, and since an objective is to minimize the variance on returns $r_i$, there is an extra constraint with respect to a targeted return $r_p$.

This problem can be turned into a new, unconstrained problem that uses Lagrange multipliers $\lambda$ and $\gamma$,
$$L[\textbf{w},\lambda,\gamma] = \frac{1}{2} \textbf{w}^T \boldsymbol\Sigma \textbf{w} - \gamma(\textbf{w}^T \textbf{1} -1) - \lambda (\textbf{w}^T \textbf{r} - r_p).$$

To find a minimum, it is possible to differentiate with respect to all the parameters of the Lagrange function and set the resulting equations equal to zero. 
This generates a system of $(N + 2)$ linear equations which can be solved to find the resulting $\textbf{w}$ weight vector of allocations.

While the approach is the seminal solution for portfolio allocation, it is recognized that it has some problems related to instability, concentration and under-performance \cite{Elton_2014a}. An open source implementation of this approach is available in \cite{a6010169}.

\subsection{Inverse Variance Portfolio (IVP)}
The rationale behind the Inverse Volatility Portfolio (IVP) allocation strategy is, in essence, simple: investments on assets are weighted in proportion to the inverse of the assets’ volatilities. 
Thus, given the standard deviation of returns $\boldsymbol\sigma$, IVP assigns the following weights to the N assets \cite{SHIMIZU2020101438}
$$\mathbf{w}= \frac{\boldsymbol\sigma^{-1}}{\mathbf{1}^T \boldsymbol\sigma^{-1}}$$
in matrix form or, alternatively, the weight $w_i$ to be assigned to each asset $a_i$ is measured as
$$w_i = \frac{1/\sigma_i}{\sum_{k=1}^N 1/\sigma_k }.$$

\subsection{Hierarchical Risk Parity (HRP)}
The Hierarchical risk parity (HRP) is an approach based on the use of graph theory and machine learning to build a diversified portfolio \cite{deprado}. An overview of the approach is reported in Algorithm \ref{alg:hrp}.
The algorithm operates in three stages. The first step uses a hierarchical clustering scheme that clusters similar assets based on their correlation. 
To perform the clustering, a notion of distance among assets is defined, that is based on their correlation level. In particular, the defined distance between two assets $a_i,a_j$ is defined as $d_{ij} = \sqrt{\frac{1}{2}(1-\rho_{ij})}$, where $\rho_{ij}$ is the correlation between $a_i$ and $a_j$.
Based on such distance metrics, near assets are combined into the same clusters.

The second step consists in rearranging rows and columns of the covariance matrix, so as to obtain a quasi-diagonal covariance matrix. The rationale is to have a matrix with high correlations placed close to each other. Quasi-diagonalization ensures that similar investments are grouped together and dissimilar ones are kept fairly apart. 

The third step consists in providing weights to assets, which are distributed using an inverse-variance allocation scheme. This is accomplished by recursively bisecting the rearranged covariance matrix. In particular, in the original paper an IVP approach is applied to assets within a cluster, but alternatives are possible \cite{deprado,doi:https://doi.org/10.1002/9781119751182.ch9}.
An implementation of this scheme is available in \cite{Martin2021}.

\begin{algorithm}
    \caption{HPR Allocation Scheme}
    \label{alg:hrp}
    \SetAlgoLined
    \DontPrintSemicolon
    \KwIn{\textit{corr}: Correlation matrix}
    \KwResult{$w$: allocation weights}
    $d \leftarrow \sqrt{\frac{1}{2}(1-\textit{corr})}$\;
    $\textit{links} \leftarrow$ clustering(\textit{d})\;
    $\textit{sortedId} \leftarrow$ quasiDiagonalization(\textit{links})\;
    $\textit{w} \leftarrow$ recursiveBisectionPartition(\textit{sortedId})\;
    \KwRet \textit{w}\;
\end{algorithm}

This approach is a seminal scheme on the use of clustering techniques applied to portfolio allocation. However, some possible limitations have been discussed in the literature. In particular, the use of the ``single linkage'', as the hierarchical clustering technique, can create a tree of similar assets which might be very deep and wide. This might prevent the creation of dense clusters and affect the weight allocations \cite{Papenbrock2011_1000025469}. Indeed, large weights can be allocated to few assets, with a resulting unequal distribution of the portfolio.

\subsection{Naive Network Modularity based Allocation (NetMod)}\label{sec:netmod}
This approach is based on the idea that a set of assets can be considered as a set of entities that share some characteristics and trends. Thus, assets form a complex network where nodes (assets) can be linked together based on their level of similarity. 
Complex network theory is an area of scientific research that is based on the idea to see everything as a network. 
This idea is largely inspired by empirical findings that extract meaningful mathematical properties of real-world networks, such as computer networks, IoT and technological systems, social networks, as well as biological and climate ones, gossip and epidemic schemes \cite{Blondel_2008,Ferretti20131631,Ferretti2017271,Newman03thestructure}.

In this work, we keep a naive approach based on the correlation matrix, i.e., in the network, nodes are the assets and a link $(i,j)$ exists between nodes $i,j$ if they have a correlation $\rho_{ij} > \alpha$, where $\alpha$ is a parameter \cite{lohre2014use}. Links $(i,j)$ have a weight $A_{ij}$ equal to $A_{ij}=\rho_{ij}$.\footnote{Notice that, here, we refer to the weight of the link in the graph, not to the weights of the portfolio allocation, used to distribute the investments over multiple assets.}

Given the network, 
what we do next is to try to identify communities of assets that are similar.
To this aim, we apply the Louvain algorithm to measure the network modularity and extract communities of assets \cite{newmancommunity}. 
Modularity is a metrics that measures the strength of a division of a network into different clusters (communities). Modularity is usually comprised between $-0.5$ and $1$, i.e., when equal to $-0.5$, it is not possible to find a good partitioning of the net into communities; when equal to $1$, a good partitioning is possible, with dense connections between the nodes within communities but sparse connections between nodes in different communities. 

The formula to measure modularity is \cite{Blondel_2008}
$$Q=\frac{1}{2m}\sum_{ij}\Big(A_{ij}-\frac{k_ik_j}{2m}\Big)\delta(c_i,c_j)$$
where $A_{ij}$ represents the link weight between nodes $i$ and $j$;
$k_{i}$ is the weighted degree to node $i$ (same is for $j$); $m$ is the sum of all of the link weights in the net; $c_{i}$ represents the community of $i$; $\delta(x,y)$ is the Kronecker delta function.

The Louvain algorithm operates in a greedy manner, by iteratively repeating two steps \cite{Blondel_2008}. The first step is devoted to find small communities, on a local basis. Thus, it starts with each node that is assigned to a different community. Then, for each node $i$ and each of its neighbours $j$, it is checked if the overall modularity increases by grouping $i$ and $j$. 
In the second step, the identified communities are grouped into a single node to pass through another iteration. 
After each iteration, the number of thus communities reduces, and the iteration continues until no changes are observed.

In NetMod, once communities of assets have been identified, a simple asset allocation scheme is employed, where the weight to be allocated is equally partitioned for each community, and this amount is once again equally distributed among assets in the community (see Algorithm \ref{algo}).

\begin{algorithm}
    \caption{NetMod Allocation Scheme}
    \label{algo}
    \SetAlgoLined
    \DontPrintSemicolon
    \KwIn{\textit{corr}: Correlation matrix, $\alpha$: Threshold for link creation}
    \KwResult{$w$: allocation weights}
    $\textit{g} \leftarrow$ createCorrelationNet(\textit{corr}, $\alpha$)\;
    $\textit{partition} \leftarrow$ communityLouvain(\textit{g})\;
    $\textit{weightPerCluster} \leftarrow$ 1/$|\textit{partition}|$\;
    \For{\textit{clus} in \textit{partition}}{
        \For{\textit{i} in \textit{clus}}{
            $w_i \leftarrow \textit{weightPerCluster}/|\textit{clus}|$\;
        }
    }
    \KwRet \textit{w}\;
\end{algorithm}

Clearly enough, this is a simple approach, which has been devised as a basic strategy to understand if the notion of modularity brings some interesting outcomes. We claim that, starting from this proposal, further research might lead to novel optimized portfolio allocation schemes. An open source implementation of the described scheme, in python code, is available in \cite{git-netmod}.


\section{An evaluation study of different portfolio allocation strategies}
\label{sec:eval}

In this section, we discuss the main building blocks used to perform the evaluation analysis of the compared portfolio allocation schemes. Thus, we present the types of simulations that allow generating different data traces for assets prices and returns. Then, we discuss the main metrics of interest to assess how the portfolio allocation strategies perform with different types of data traces.

\subsection{Monte-Carlo simulation framework}
To assess all the considered portfolio allocation methods in multiple and heterogeneous situations, different types of simulations have been implemented, that generate synthetic return traces with diverse characteristics. 
Return $r_t$ of an asset at time $t$ is obtained through the price $p_t$ of the asset in different moments, i.e.~$r_t=\frac{p_t-p_{t-1}}{p_{t-1}}$.

Algorithm \ref{alg:mcsim} shows a simplified sketch of the Monte-Carlo simulation to clarify how data are generated and used to assess the portfolio simulation scheme. 
The parameters that define the simulation environment are the size of the corpus of simulations (\textit{numIters}, i.e., how many different simulation runs are performed for the considered scenario), the length of the data trace (\textit{simLength}, i.e., the duration of each simulation run, in terms of number of assets' prices and related returns observations), the size of the time interval after which the weights of the portfolio allocation are tuned and rebalanced based on the return observations (\textit{deltaT}, i.e., how often the portfolio is updated), the number of assets to be generated (\textit{numAssets}), the data trace generation type (\textit{dataType}, i.e., which scheme is employed to generate the data trace), the specific portfolio allocation scheme used to perform the allocation (\textit{allocationStrategy}).
Thus, each simulation run corresponds to a randomly generated instance of occurrences. A data trace is generated using a specific method (see below for those that have been tested during this evaluation). The data trace corresponds to a series of returns for each simulated asset. 
Then, thanks to the specific portfolio allocation scheme and based on the previous assets' return observations, the simulation computes, on a regular basis (i.e., every \textit{deltaT} timesteps in Algorithm \ref{alg:mcsim}), the weights to be associated with the set of assets. Thus, the weights are used on a period which is different from the one exploited to optimize the strategy (out-of-sample).
Through the computed weights and the returns of the assets, it is possible to measure the earned portfolio return.
During the simulation, all these results are logged to collect final statistics at the end of the execution of the whole corpus.

\begin{algorithm}
    \caption{Monte-Carlo simulation framework}
    \label{alg:mcsim}
    \SetAlgoLined
    \DontPrintSemicolon
    \KwIn{\textit{numIters}: number of simulation iterations}
    \KwIn{\textit{simLength}: length of the data trace}
    \KwIn{\textit{deltaT}: time interval for the tuning of weights}
    \KwIn{\textit{numAssets}: number of assets}    \KwIn{\textit{dataType}: pointer to the data trace generation function}
    \KwIn{\textit{allocationStrategy}: pointer to the function that allocate weights}
    
    \For{\textit{i} in \textit{numIters}}{
        \textit{trace} $\leftarrow$ generateData(\textit{dataType}, \textit{simLength}, \textit{numAssets})\;
        \textit{ints} = divideIntervals(\textit{trace}, \textit{deltaT})\;
        \For{$x$ in \textit{ints}}{
          \textit{corr} = measureCorrelation(trace)\;
          $w$ = applyScheme(\textit{allocationStrategy}, \textit{corr})\;
          $r = x\cdot w$\;
        }
        collectStats()\;
    }
    makeStatistics()\;
\end{algorithm}

\begin{figure*}
    \centering
    \includegraphics[width=.4\linewidth]{"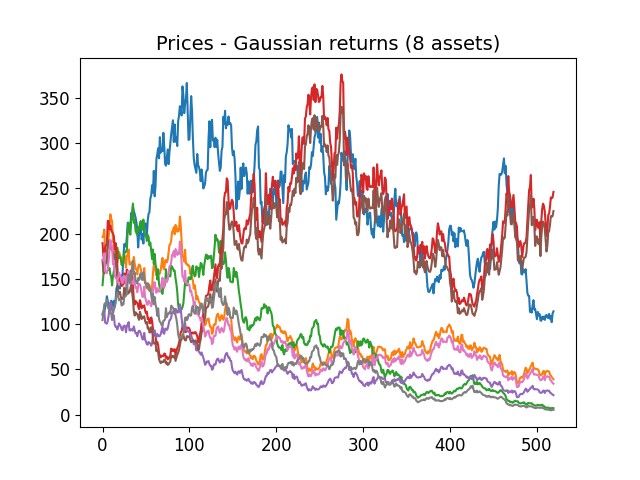"}
    \includegraphics[width=.4\linewidth]{"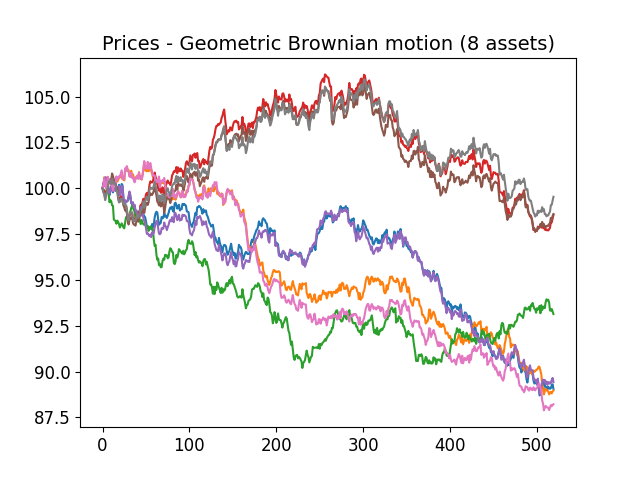"}
    \includegraphics[width=.4\linewidth]{"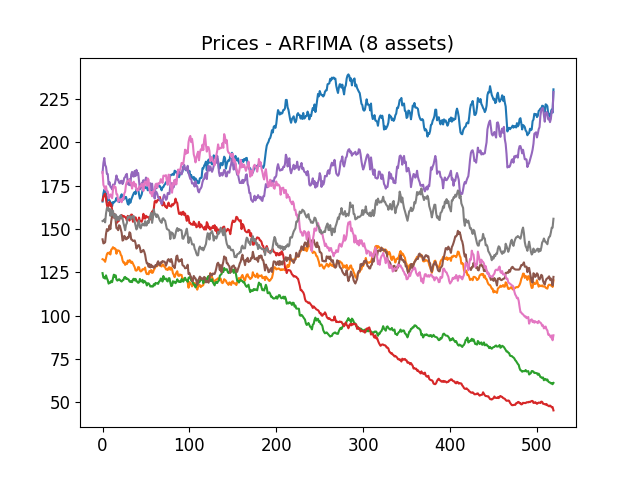"}
    \includegraphics[width=.4\linewidth]{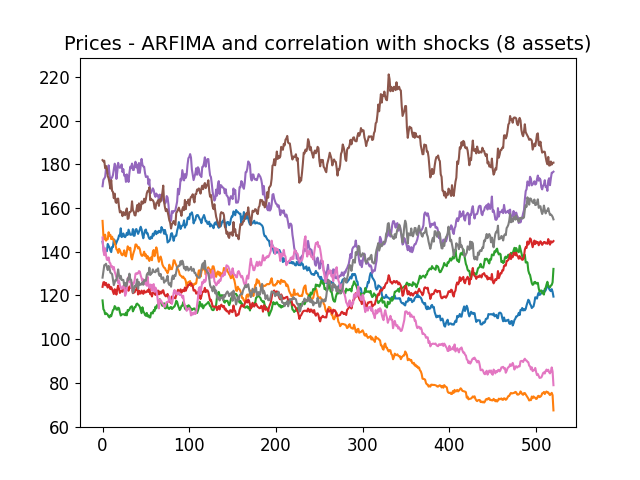}
    \includegraphics[width=.4\linewidth]{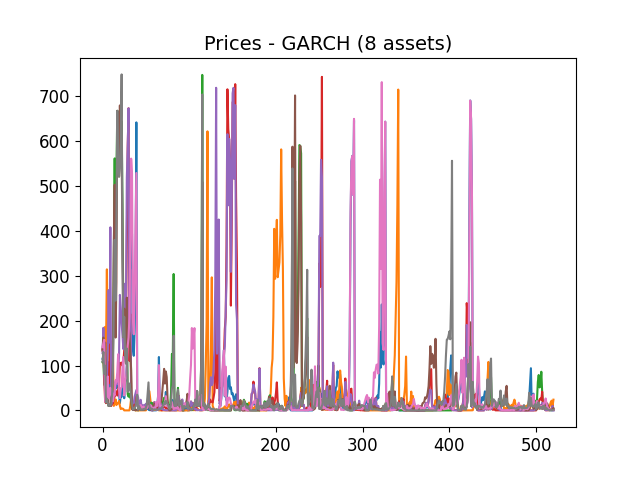}
    \caption{Exemplars of different data traces depending on the employed simulations - $8$ assets.}
    \label{fig:traces}
\end{figure*}

The implemented types of simulation scenarios are discussed in the following. 
Figure \ref{fig:traces} shows examples of possible sets of generated traces, when the number of synthetic generated assets is equal to $8$. 
The figure should clarify that each simulation type provides a diverse scenario where the portfolio allocation methods are employed and analyzed. This provides a wider and more general understanding of the performances of all these methods, rather than using classic stationary data traces, only.

In the simulation tests discussed in the rest of the paper, the following parameter setting was used. We repeated an amount of $100$ simulation runs for each simulation corpus (i.e., each type of simulation). The length of the generated data trace was equal to $520$ (equivalent to two years of daily history), while the \textit{deltaT} parameter was set equal to $60$. The number of simulated assets was equal to $32$. The window period employed to measure the values of DCCA and DPCCA (i.e., the parameter $n$ in Section \ref{sec:dcca}) was set equal to $60$.

\subsubsection{Gaussian Returns simulations}
We already mentioned that most studies on portfolio allocation employ the assumption that stationary data series can be used in their analyses \cite{10.1080/713665670}. As a matter fact, assets' price time series are commonly non-stationary \cite{RePEc:eee:phsmap:v:392:y:2013:i:8:p:1756-1761}. Thus, to perform inferential analyses, researchers usually focus on returns on prices, assuming that these can be better approximated via stationary processes. 
Following this idea, returns are often modeled via Normal distributions \cite{deprado}. 

This type of simulation follows such an approach, i.e., assets returns are generated using Gaussian distributions. 
In particular, $x$ asset returns are generated using independent Guassian distributions. Then, other $y$ asset returns are generated to be dependent on the previous ones. Each of these $y$ dependent traces is generated by randomly taking one of the independent $x$ assets' returns, and adding some white noise, i.e., another (steeper) Gaussian distribution.
In Section \ref{sec:res}, we will show results from a corpus of simulations where the number of assets was equal to $32$, with $x=16, y=16$.

While this represents a standard approach to perform Monte-Carlo simulations in financial applications, it has been shown that financial asset return distributions are usually not Normal \cite{10.1080/713665670}. For this reason, other simulation types for the generation of assets returns are considered in the following.

\subsubsection{Geometric Brownian motion}
The Geometric Brownian Motion (GBM) simulation scheme implements a continuous-time stochastic process, in which the logarithm of the randomly varying quantity of interest follows a Brownian motion with drift \cite{doi:10.1119/1.16497}.
GBM is a popular scheme, that has been used to underlie the dynamics of a diverse set of natural phenomena, including finance, distribution of incomes, body weights, weather forecasts, fragment sizes in rock crushing processes \cite{MERTON1976125}. It is quite useful to model non-stationary random processes. This is obtained by introducing a stochastic drift on the random process.

More in detail, the process is based on a stochastic differential equation
$$dr_t=\mu r_t dt + \sigma r_t dW_t,$$
where  $W_t$ is a Wiener process, $\mu$ represents the stochastic drift and $\sigma$ is a measure of the volatility.
A Weiner process $W_t$ is a stochastic process characterized by these properties: 
i) $W_0=0$; ii) the increments $W_{t+u} - W_t, u\geq 0$, are independent random variables; put in other words, the increment from $W_t$ to $W_{t+u}$ does not depend on the values of the process before $t$ ($W_s, s \leq t$); iii) such increments $W_{t+u} - W_t$ are Normally distributed with mean $0$ and variance $u$, i.e., $W_{t+u}-W_t \sim \mathcal{N}(0,u)$ \cite{KaratzasIoannis1998Bmas}.

The solution to the differential equation above is 
$$r_t= r_0 e^{(\mu -(\sigma^2/2))t + \sigma W_t}$$
that is the formula used to generate the data traces in these simulations.
Also in this case, in Section \ref{sec:res} we will show results from a corpus of simulations where the number of assets was equal to $32$, with $x=16, y=16$.

\subsubsection{GARCH}
The Generalized Auto Regressive Conditional Heteroskedasticity (GARCH) model is a type of Auto Regressive Moving Average (ARMA) model, applied to the variance of the time series. It is composed of an auto-regressive term and a moving average term.
The rationale around the use of this model in finance is due to the basic observation that, typically, large asset returns tend to be followed by more large returns \cite{9400}.
Thus, the volatility of asset returns is usually serially correlated.
The GARCH($p$,$q$) model is characterized by two parameters $p$, $q$. The variance of the time series is characterized as follows
$$\sigma_t^2=\alpha_0 + \sum_{i=1}^p \alpha_i \varepsilon_{t-i}^2 + \sum_{j=1}^q \beta_j \sigma_{t-j}^2.$$
In order for $\sigma_t^2>0$, it is assumed that $\alpha_0>0$ and the other coefficients $\alpha_i, \beta_j$ are all non-negative. Usually, the GARCH(1,1) model is employed to model the volatility of daily returns, which are in turn calculated based $\sigma_t$ values \cite{9400},
\begin{align*}
r_t & = \varepsilon_t = \sigma_t z_t, \ z_t \sim N(0,1)\\
\sigma_t^2 & = \alpha_0 + \alpha_1 \varepsilon_{t-1}^2 + \beta_1 \sigma_{t-1}^2.
\end{align*}
Here, $z_t$ is white noise. Moreover, we should keep $\alpha_1 + \beta_1 <1$ to avoid that the model is unstable.

As per other types of simulation, $x$ traces are created. Then, $y$ traces are generated by randomly taking one of the independent $x$ ones, and adding some white noise based on a Gaussian distribution (in the experiments, Section \ref{sec:res}, the amount of traces was $x=y=16$).

\subsubsection{ARFIMA}
The AutoRegressive Fractionally Integrated Moving Average (ARFIMA) process was introduced to generate time series with power-law correlations \cite{doi.org/10.1111/j.1467-9892.1980.tb00297.x}.
It has been recognized that ARFIMA models, when applied to financial time series, sometimes provide significantly better out-of-sample data traces than AR, MA, ARMA, GARCH, and related models \cite{RePEc:rut:rutres:200422}.

In particular, according to the ARFIMA model each generated variable (asset return) depends not only on its own past, but also on the past values of the other variables.
We start by creating pairs of times series $r_1, r_2$ with long range cross-correlations \cite{RePEc:eee:phsmap:v:392:y:2013:i:8:p:1756-1761},
$$r_{1,i} = W \sum_{n=1}^\infty a_n(\rho_1)r_{1,i-n} + (1-W) \sum_{n=1}^\infty a_n(\rho_2)r_{2,i-n} + \varepsilon_{1,i},$$
$$r_{2,i} = (1-W)\sum_{n=1}^\infty a_n(\rho_1)r_{1,i-n} + W \sum_{n=1}^\infty a_n(\rho_2)r_{2,i-n} + \varepsilon_{2,i},$$
where $W$ is a weight value ($W \in [0.5,1]$) that controls the strength of the correlations between the two traces $r_1$ and $r_2$;
$\varepsilon_{1,i}$ and $\varepsilon_{2,i}$ are independent and identically distributed Gaussian variables with zero mean and unit variance, i.e.~$\mathcal{N}(0,1)$, representing white noise; $a_n(\rho)$ are statistical weights defined as 
$$a_n(\rho) = \Gamma (n-\rho)/(\Gamma(-\rho)\Gamma(1 + n)),$$ 
being $\Gamma$ the Gamma function; the $\rho$ parameters range in the interval $\rho \in [-0.5,0.5]$.

Following this definition, in this type of simulation, we thus created a set of pairs of $x$ ARFIMA traces (as before, $x=16$ in Section \ref{sec:res}). Then, an additional set of other correlated traces was created, similarly to the previous case. Thus, other $y$ traces were generated by randomly taking one of the independent $x$ ones, and adding some white noise based on a Gaussian distribution ($y=16$, in Section \ref{sec:res}).

\subsubsection{ARFIMA and correlation with shocks}
In this case, we created a combination of traces generated through ARFIMA processes, as described above.
After this previous phase, some pairs of asset returns, say $a$ and $b$, were stochastically modified, in randomly chosen time intervals, by changing their values as follows: $r_{a,i}=\beta r_{a,i} + (1-\beta)r_{b,i}$, $r_{b,i}=\beta r_{b,i} + (1-\beta)r_{a,i}$. This was motivated by the idea of adding correlation to certain assets, during limited time intervals. The rationale was to try to understand if sporadic higher correlations are captured by the considered portfolio allocation strategies.

Finally, 
we added a random amount of shocks to some randomly selected data traces. In particular, we generate a random amount of shocks. For each shock, a random asset $a$ was chosen, as well as a random point in time $t_i$, i.e.~the time when the shock started. A random duration of the shock $d$ was generated, comprised in the range $[1, T/10]$ (being $T$ the length of the data trace). For the randomly computed time period $[t_i, t_{i+d}]$, the values of the returns $r_{a,j}$ of asset $a$ at time $t_j$ were updated as follows:
$$r_{a,j} = r_{a,j} + \alpha\ \text{ran}(-r_a^{max}, r_a^{max}),\ \ j = i, \dots, i+d$$
being $\alpha$ a randomly chosen value in $[0,1)$ set for the whole shock interval, and $\text{ran}(-r_a^{max}, r_a^{max})$ a uniformly random generated value in an interval regulated by the highest return for asset $a$, i.e., $r_a^{max}$.

\subsection{Backtest}
As already mentioned, wide simulation studies are more profitable to study how a given method can perform in general, rather than focusing on some given historical data traces. However, as further validation, some backtests have been performed. In particular, data traces of a two years time interval of a set of assets were collected from Yahoo finance. Then, a backtest simulation was accomplished over these historical data traces.

\begin{figure*}
\centering
  \includegraphics[width=.6\linewidth]{"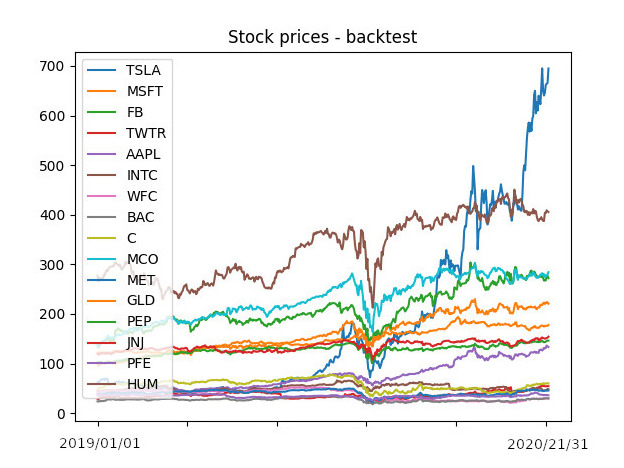"}
\caption{Backtest - asset prices}
\label{fig:bt-traces}
\end{figure*}

Indeed, in order to verify how these studied schemes might perform in a real scenario, we took prices of a set of real assets in the period 2019/01/01 - 2020/12/31. The considered assets were: IT assets, i.e.~Tesla (TSLA), Microsoft (MSFT), Facebook (FB), Twitter (TWTR), Apple (AAPL), Intel (INTC); financial assets, i.e.~Wells Fargo (WFC), Bank of America Corp.~(BAC), Citigroup Inc.~(C), Moody's Corp.~(MCO), MetLife Inc.~(MET); general assets or stocks, i.e.~SPDR Gold Shares (GLD), PepsiCo, Inc.~(PEP); healthcare assets, i.e.~Johnson \& Johnson (JNJ), Pfizer Inc.~(PFE), Humana Inc.~(HUM). 
The rationale behind the choice of these assets was to select them from different sectors, assuming that a certain level of correlation may exist among assets of the same sector. 
The length of the time series was in line with those generated in the simulation study.
Figure \ref{fig:bt-traces} shows the related stock prices.

\subsection{Considered Metrics}
In this section, we describe the main metrics used to evaluate the performance of different portfolio allocation schemes.
We will report statistical measures related to the returns obtained by applying the different allocation schemes.
Moreover, compound log returns are measured, which are obtained by summing log returns. It is a quite commonly exploited measure in finance, since log returns usually show a reduced variation on the time series, thus making it easier to fit the models, when needed. Furthermore, the following metrics have been measured.

\subsubsection{Portfolio Variance}
Portfolio variance of a portfolio allocation, at time $t$, is measured as 
$$\text{PV}_t(\textbf{w}_t,\boldsymbol\Sigma_t) = \textbf{w}_t^T \boldsymbol\Sigma_t \textbf{w}_t,$$
where $\textbf{w}_t$ is the vector of weights associated with the set of assets at time $t$, $\boldsymbol\Sigma_t$ is the covariance matrix of returns (at time $t$), and $\textbf{w}_t^T$ is the transpose of $\textbf{w}_t$.
Higher portfolio variance corresponds to higher risks, and thus it is considered as an indicator of bad performance.

\subsubsection{Risk Contribution}
The risk contribution of a given asset $j$ to the total portfolio can be measured as
$$\text{RC}_j(\textbf{w},\boldsymbol\Sigma) = w_j\frac{(\boldsymbol\Sigma \textbf{w})_j}{\sqrt{\textbf{w}^T \boldsymbol\Sigma \textbf{w}}},$$
where $w_j$ is the weight associated with asset $a_j$ and $(\boldsymbol\Sigma \textbf{w})_j$ is the j-th component of $\boldsymbol\Sigma \textbf{w}$. 
(Note that for the sake of readability, we removed the time $t$ subscripts.)

Risk contribution is the ratio between a weighted average of asset volatility and portfolio volatility.
Comparing the assets' risk contribution, obtained through the use of different allocation methods, provides an estimation of portfolio diversification.
It is thus interesting to measure the distribution of risks associated with assets, based on their weights. 

\subsubsection{Diversification Ratio}
The Diversification Ratio (DR) of a given portfolio allocation of weights at time $t$, $w_t$, is computed through the following equation:
$$\text{DR}(\textbf{w}_t,\boldsymbol\Sigma_t)=\frac{\textbf{w}_t^T \boldsymbol\Sigma_t}{\sqrt{\textbf{w}_t^T \boldsymbol\Sigma_t \textbf{w}_t}}.$$ The diversification ratio is the ratio of the weighted average of volatilities divided by the portfolio volatility (square of the portfolio variance) \cite{Choueifaty40}. 
It is similar to the risk contribution above, but in this case, it is a global metrics that gives the measure of diversification in the portfolio. A higher diversification ratio is a better performance indicator.

\subsubsection{Normalized HHI}
Herfindahl-Hirschman Index (HHI) is a common measure of market concentration and is used to determine market competitiveness \cite{10.2307/136100}.
It is defined as
$$\text{HHI}(\mathbf{w})=\sum_{i=1}^N w_i^2,$$
where N is the number of considered assets.
The Herfindahl Index (H) ranges from 1/N to one.
Actually, we will consider the Normalized HHI (NHHI), which is measured as $$\text{NHHI}(\mathbf{w}) =\frac{(\text{HHI}(\mathbf{w})-1)/N}{1-1/N}.$$
The benefit of such normalized version of the index is that it ranges from 0 to 1. Thus, information about the total number of assets (N) is lost, in favor of a more general view of such an index as a measure for the equality of distributions.

Indeed, a greater value of the index reflects greater risk concentration, while a small index indicates a competitive market, with no dominant assets. 
Thus, in our scenario, the lower the better.

\subsubsection{Sharpe Ratio}
The Sharpe Ratio (SR) is a measure of the expected return of an investment, versus the amount of variability in the return \cite{Sharpe49}. It is a very popular metrics to evaluate the performance of a portfolio allocation. The formula is 
$$\text{SR} = \frac{\hat{r}-r_{\text{free}}}{\sigma}$$
where $\hat{r}$ is the expected portfolio return, $r_{\text{free}}$ the rate you would get from a risk free investment, and $\sigma$ is the portfolio’s standard deviation. Quite often, it is commonly assumed that the risk free rate is zero, thus obtaining $\text{SR} = \frac{\hat{r}}{\sigma}$. 
Higher values of SR imply better performance of the portfolio allocation.

The main limitation of this metrics is that the SR value can be accentuated by investments that do not have a normal distribution of returns. Take for instance the case of some investment strategy that usually produces small positive returns with some occasional large negative return. By looking at historical data, one might estimate a large value of the SR, until a big loss takes place.

\subsubsection{Value at Risk and Conditional Value at Risk}
Value at Risk (VaR) is a measure of the risk of loss for investments. This measure provides a probabilistic estimation of how much a portfolio allocation will lose at worst, given normal market conditions \cite{GREGORY2008167}. There are several ways to measure VaR. Here, the calculation of the VaR is performed using historical data. 
The idea is to construct a distribution of returns based on the empirical distribution of historical returns. Thus, there is an assumption that historical returns represent the distribution of future returns.
Then, VaR is measured as the loss value $x(t)$ such that,
during the forecast horizon, it is expected that the portfolio will lose less than such $x(t)$ value, with probability $(1 -\alpha)$. 
That is, given the historical return distribution, VaR is calculated by taking the value that corresponds to the $\alpha$ percentile of such distribution.

Conditional Value at Risk (CVaR) is derived by taking a weighted average of the losses in the tail of the distribution of possible returns, beyond the VaR cutoff point. In practice, the measurement is performed by taking all the historical returns below the VaR, and calculating the mean value.

Being these metrics estimations of loss, the lower the value the better the performance of the portfolio allocation scheme.

\section{Simulation Results}
\label{sec:res}
As already mentioned, we consider four allocation schemes, i.e., CLA, IVP, HRP, NetMod. 
Besides CLA (that is naturally applied using Pearson's correlation), for other schemes we have three variants depending on the correlation metrics that we plugged into the scheme. 
Thus, in the following, all the approaches are referred to by the identifier of the scheme followed by a label that identifies the correlation metrics, e.g., ``NetMod dcca'' refers to the NetMod approach when DCCA correlation is exploited. 

For each considered scenario, we executed a corpus of $100$ simulations. Results are thus average outcomes from these simulation corpuses.
We measured the daily returns from the different simulations and performed statistical tests to understand if each scheme is significantly different from the selected baseline approach, i.e., ``IVP cov". In particular, pairwise t-tests were conducted, as well as Tukey HSD multiple comparison tests, one for each type of simulation, that consider all the schemes together.
While the obtained $p$-value of the $t$ tests varied, depending on the type of simulation and exploited approach, we never obtained a statistically significant difference ($p<0.05$) on such returns, in general.
No statistical differences were evident also from the Tukey tests.
Nevertheless, in the rest of the section we report the average daily return improvements, as well as the variation of daily returns, since in specific cases we noticed differences worthy of mention. 
Given the mean daily return $\hat{r}_x$ of a given scheme $x$, the improvement was measured as $\frac{(\hat{r}_x-\hat{r}_{\text{baseline}})}{|\hat{r}_{\text{baseline}}|} $.

Then, we also report the metrics mentioned in the previous section, i.e., Compound Log Returns (CLR), NHHI, Portfolio Variance (PV), Diversification Ratio (DR), as well as the risk contributions and average weights associated with the different assets.

As concerns those approaches exploiting DCCA and DPCCA correlations, for the sake of brevity we omit all the results related to different settings of the box length $n$, used in Equation (\ref{eq:rho}). 
In particular, the results reported in this section have been obtained with a value of $n=60$, which seemed to be a good compromise in all these different simulation scenarios. 

\subsection{Gaussian returns simulations}

\begin{table*}
\caption{Gaussian returns simulation - results}
\begin{center}
\footnotesize
\begin{tabular}{ | l | c | c | c | c | c | c | c | c | c |}
\hline
\textbf{method} &\textbf{daily ret} &\textbf{Impr} & \textbf{CLR} &  \textbf{NHHI} &\textbf{PV} & \textbf{DR} &  \textbf{SR}  &  \textbf{VaR}  &  \textbf{CVaR}\\
IVP cov & $0 \pm 0.03$ & 0 & -0.16 & 0 & 0.001 & 0.4 & 0 & 0.06 & 0.07\\
IVP dcca & $0 \pm 0.03$ & -0.37 & -0.17 & 0.01 & 0.001 & 0.37 & 0 & 0.07 & 0.08\\
IVP dpcca & $0 \pm 0.03$ & -0.37 & -0.17 & 0.01 & 0.001 & 0.37 & 0 & 0.07 & 0.08\\
HRP cov & $0 \pm 0.03$ & 1.06 & -0.1 & 0.01 & 0.001 & 0.43 & 0.01 & 0.06 & 0.07\\
HRP dcca & $0 \pm 0.03$ & -0.64 & -0.17 & 0.03 & 0.001 & 0.34 & 0.01 & 0.08 & 0.09\\
HRP dpcca & $0 \pm 0.03$ & -0.63 & -0.15 & 0.02 & 0.001 & 0.36 & 0.01 & 0.07 & 0.09\\
NetMod cov & $0 \pm 0.03$ & 1.29 & -0.06 & 0.01 & 0.001 & 0.43 & 0.01 & 0.06 & 0.07\\
NetMod dcca & $0 \pm 0.03$ & 0.27 & -0.15 & 0.01 & 0.001 & 0.37 & 0.01 & 0.07 & 0.08\\
NetMod dpcca & $0 \pm 0.03$ & 0.49 & -0.14 & 0.01 & 0.001 & 0.37 & 0.01 & 0.07 & 0.08\\
CLA cov & $0 \pm 0.06$ & -2.39 & -0.85 & 0.32 & 0.003 & 0.23 & 0 & 0.14 & 0.17\\
\hline
\end{tabular}
\end{center}\label{tab:steady}\end{table*}

\begin{figure*}
\centering
  \includegraphics[width=\linewidth]{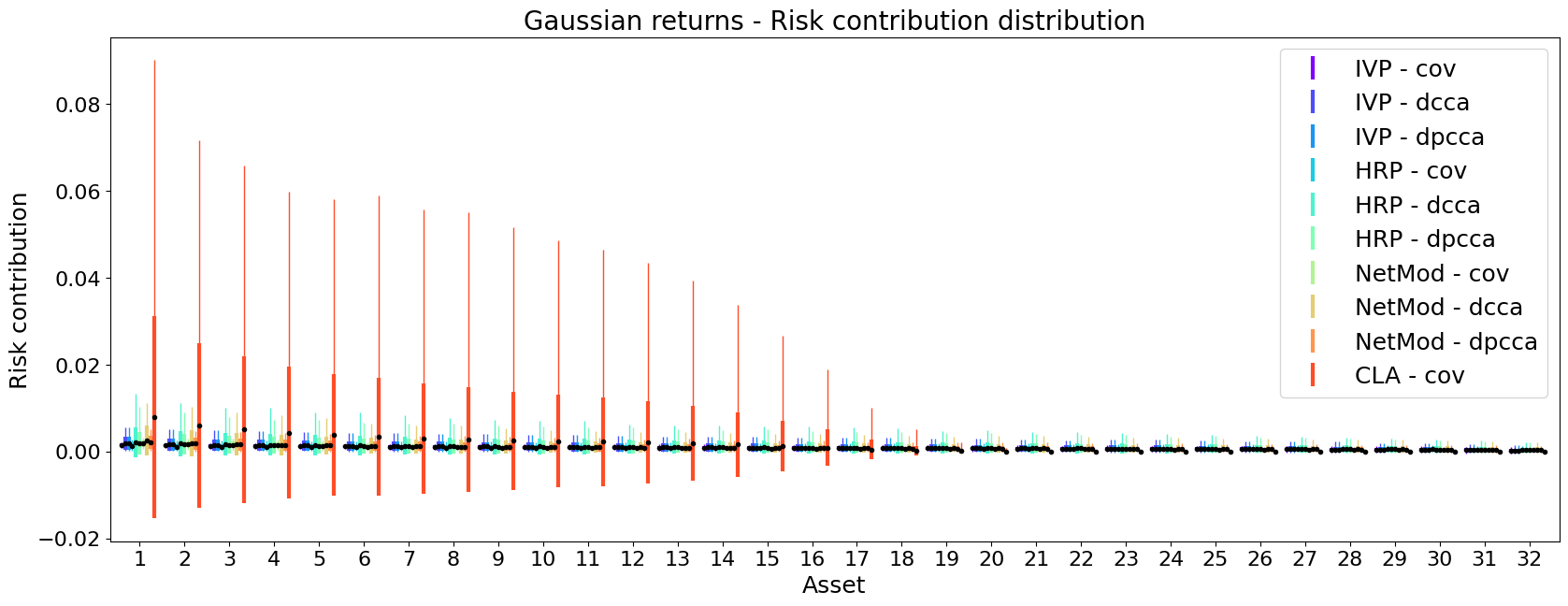}
\caption{Gaussian returns simulations: Risk contribution distribution}
\label{fig:steady-RC}
\end{figure*}

\begin{figure*}
\centering
  \includegraphics[width=\linewidth]{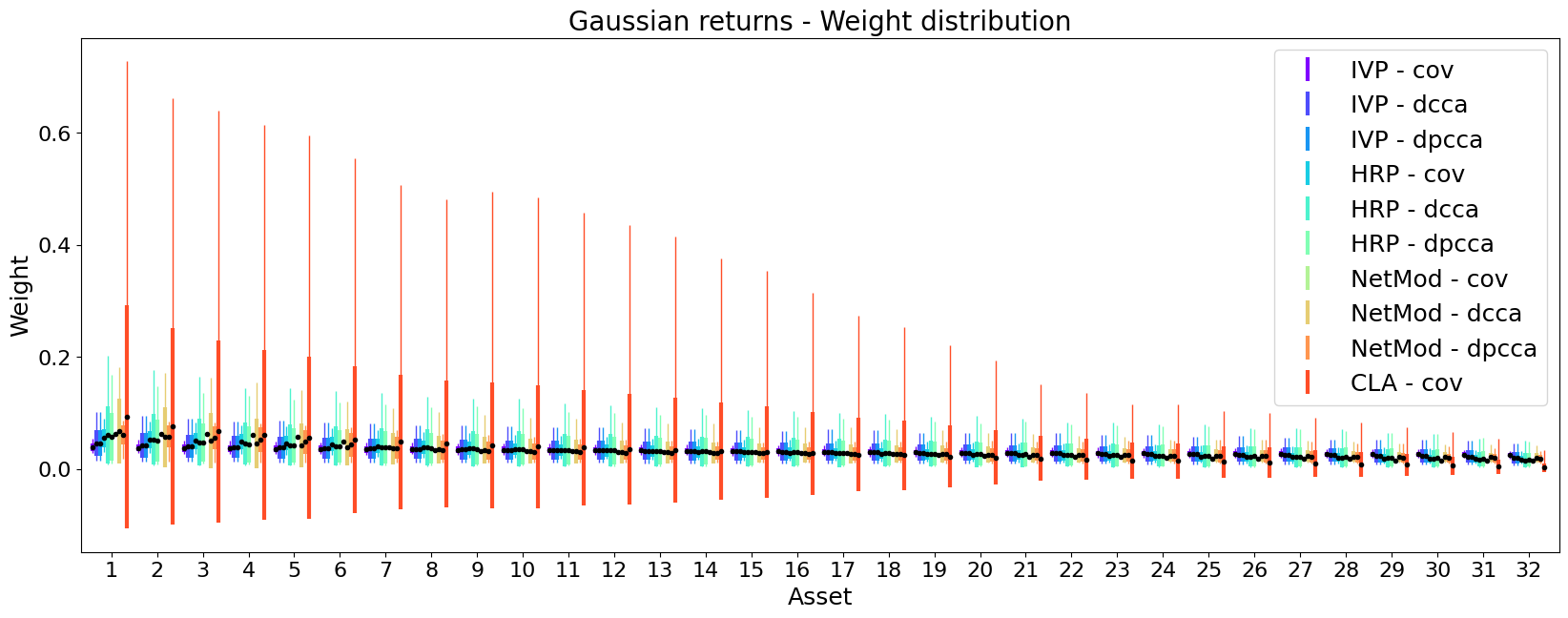} 
  \caption{Gaussian returns simulations: Weight distribution}
\label{fig:steady-weight-dist}
\end{figure*}

\begin{figure*}
\centering
  \includegraphics[width=.5\linewidth]{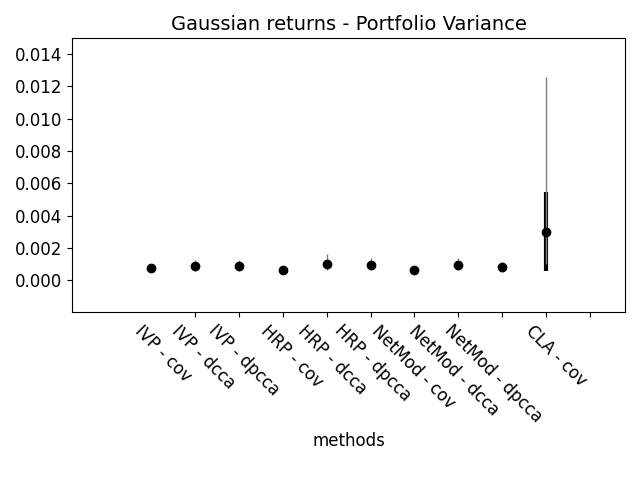}
\caption{Gaussian returns simulations: portfolio variance}
\label{fig:steady-pf}
\end{figure*}

Table \ref{tab:steady} shows the metrics of interest for the Gaussian returns simulations. 
According to these simulations, we notice that daily returns (second column of the table) varied similarly for all schemes, showing equal standard deviations, except for CLA that has a higher value.  
As concerns the average daily improvement, the best result is obtained with ``NetMod dcca". An interesting result is related to the NHHI metrics, with ``IVP cov'' that outperforms other approaches (the lower the better), followed by other IVP, NetMod approaches and ``HRP cov''. In general, these NHHI values demonstrate that all the approaches equally distribute the weights of the allocations, except for CLA that has a higher value.
All the schemes show similar average PVs (the lower the better) DRs (the higher the better), SRs (the higher the better), VaR and CVaR (the lower the better), with worse results for CLA and slightly better results for NetMod approaches. 

The outcomes outlined in the previous table are confirmed by the next figures.
Figure \ref{fig:steady-RC} shows the risk contribution distribution on different assets, depending on the employed allocation scheme and correlation metrics.
In the chart, assets are ordered according to the measured risk contribution.
For each asset, we show the mean value (central dot in the bar), the standard deviation (thicker vertical bar), min and max values (thinner vertical bar). 
This figure provides a general overview on how each method distributes the risks to different assets.
A general outcome is that CLA has a wider variance and under-utilizes the available assets. Among the others, HRP methods tend to have wider variances, followed by NetMod schemes and finally IVP ones. In general, for each allocation scheme we notice that ``dcca" and ``dpcca" variants tend to have slightly wider standard deviation, than ``cov" (probably not visible in the figure).
These results are confirmed by Figure \ref{fig:steady-weight-dist}, where similar statistics are shown on the weight distribution (ordered according to higher to lower weights).
Also in this case, it is confirmed that CLA has a wider standard deviation and in general ``dcca" and ``dpcca" variants have slightly higher values than ``cov".

The mentioned results on the volatility of the allocations are confirmed from the average values of the PVs obtained according to the different portfolio allocation methods, which are shown in Figure \ref{fig:steady-pf}. In this case, it is possible to observe how CLA is worse than other approaches.

\subsection{Geometric Brownian motion simulations}

\begin{table*}
\caption{Geometric Brownian motion simulation - results}
\begin{center}
\footnotesize
\begin{tabular}{ | l | c | c | c | c | c | c | c | c | c |}
\hline
\textbf{method} &\textbf{daily ret} &\textbf{Impr} & \textbf{CLR} &  \textbf{NHHI} &\textbf{PV} & \textbf{DR} &  \textbf{SR}  &  \textbf{VaR}  &  \textbf{CVaR}\\
IVP cov & $0 \pm 0$ & 0 & -0.23 & 0.03 & 0 & 0.03 & -0.45 & 0 & 0\\
IVP dcca & $0 \pm 0$ & 0.01 & -0.23 & 0.05 & 0 & 0.03 & -0.4 & 0 & 0\\
IVP dpcca & $0 \pm 0$ & 0.01 & -0.23 & 0.05 & 0 & 0.03 & -0.4 & 0 & 0\\
HRP cov & $0 \pm 0$ & 0 & -0.23 & 0.04 & 0 & 0.04 & -0.44 & 0 & 0\\
HRP dcca & $0 \pm 0$ & 0.01 & -0.23 & 0.08 & 0 & 0.03 & -0.35 & 0 & 0\\
HRP dpcca & $0 \pm 0$ & 0.02 & -0.22 & 0.06 & 0 & 0.03 & -0.37 & 0 & 0\\
NetMod cov & $0 \pm 0.01$ & 0.05 & -0.23 & 0 & 0 & 0.2 & -0.07 & 0.02 & 0.02\\
NetMod dcca & $0 \pm 0.01$ & 0.01 & -0.24 & 0.02 & 0 & 0.18 & -0.06 & 0.02 & 0.02\\
NetMod dpcca & $0 \pm 0.01$ & 0.07 & -0.23 & 0.03 & 0 & 0.16 & -0.05 & 0.02 & 0.03\\
CLA cov & $0 \pm 0$ & 0.01 & -0.23 & 0.23 & 0 & 0.08 & -0.12 & 0.01 & 0.01\\
\hline
\end{tabular}
\end{center}\label{tab:gbm}
\end{table*}

\begin{figure*}
\centering
  \includegraphics[width=\linewidth]{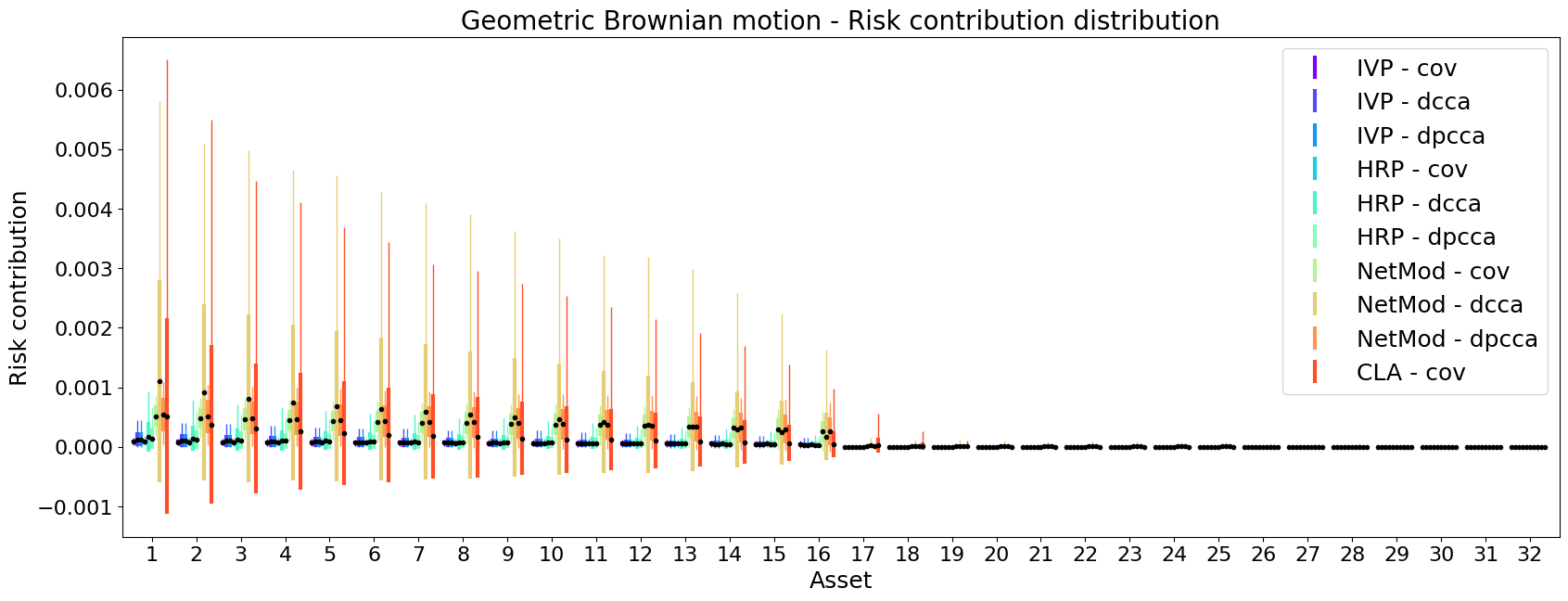}
\caption{Geometric Brownian motion: Risk contribution distribution}
\label{fig:gbm-RC}
\end{figure*}

\begin{figure*}
\centering
  \includegraphics[width=\linewidth]{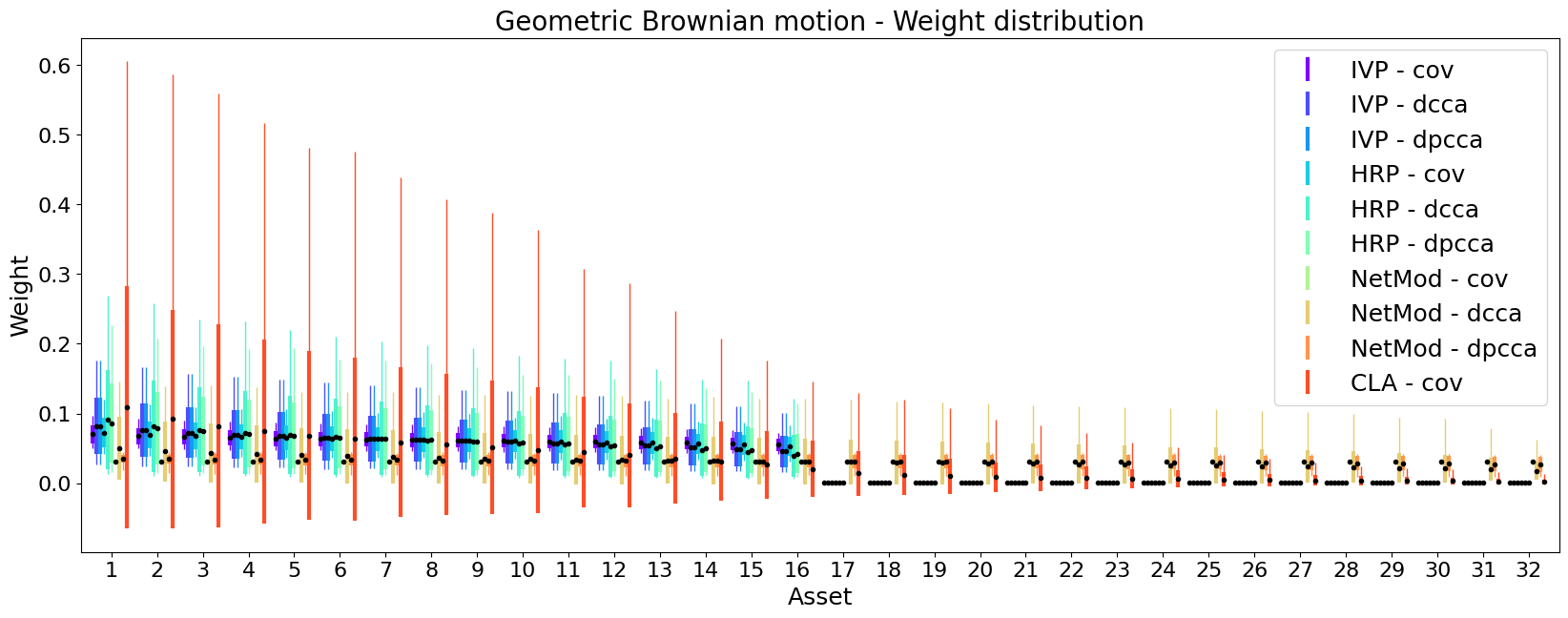}
\caption{Geometric Brownian motion: Weight distribution}
\label{fig:gbm-weight-dist}
\end{figure*}

\begin{figure*}
\centering
  \includegraphics[width=.5\linewidth]{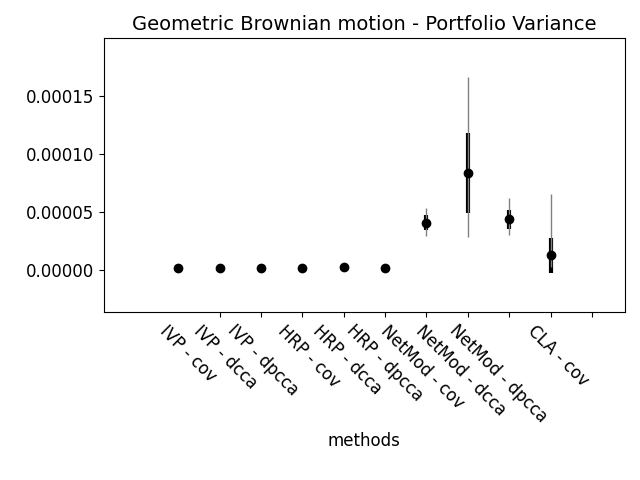}
\caption{Geometric Brownian motion: portfolio variance}
\label{fig:gbm-pf}
\end{figure*}

Table \ref{tab:gbm} shows the metrics obtained for the Geometric Brownian motion simulations, similarly to what has been done in the previous section. 
The standard deviation of daily returns slightly varies for all the schemes. NetMod methods seem to be slightly more risky (higher standard deviations, VaR and CVaR), yet offering higher daily returns. Indeed, average PVs are all negligible. While results on the standard deviation penalize NetMod schemes, however, these approaches show higher (better) DRs. 
This result is confirmed from the NHHI measurements, which are lower (better) for NetMod schemes, especially ``NetMod cov". NetMod outperforms others also in terms of SR.

An interesting outcome is evident from Figures \ref{fig:gbm-RC}--\ref{fig:gbm-weight-dist}, where the risk distributions and weight distribution allocations are shown for the different assets (ordered in decreasing order). All the employed methods, except for ``NetMod" ones, do not allocate positive weights (and thus no risk) to half of the assets. As mentioned, the assets have been generated to be highly correlated i.e., half of the assets are independent, while others are generated starting with a given randomly chosen asset and by adding some random noise.
Thus, these approaches select one among the highly correlated assets ad avoid the use of the other ones. Conversely, by design the ``NetMod" schemes identify such correlation, but then equally distribute the portion of the weights to be allocated to each specific cluster of correlated assets.
This results in a more varied use of the considered assets and in a larger (yet limited in value) portfolio variance, as shown in Figure \ref{fig:gbm-pf}.

\subsection{GARCH simulations}

\begin{table*}
\caption{GARCH simulation - results}
\begin{center}
\footnotesize
\begin{tabular}{ | l | c | c | c | c | c | c | c | c | c |}
\hline
\textbf{method} &\textbf{daily ret} &\textbf{Impr} & \textbf{CLR} &  \textbf{NHHI} &\textbf{PV} & \textbf{DR} &  \textbf{SR}  &  \textbf{VaR}  &  \textbf{CVaR}\\
IVP cov & $0.09 \pm 0.24$ & 0 & 26.87 & 0 & 0.044 & 3.14 & 0.36 & 0.4 & 0.45\\
IVP dcca & $0.09 \pm 0.26$ & -0.01 & 24.19 & 0.02 & 0.052 & 2.82 & 0.33 & 0.44 & 0.49\\
IVP dpcca & $0.09 \pm 0.26$ & -0.01 & 24.19 & 0.02 & 0.052 & 2.82 & 0.33 & 0.44 & 0.49\\
HRP cov & $0.09 \pm 0.23$ & -0.01 & 27.13 & 0.01 & 0.037 & 3.31 & 0.37 & 0.39 & 0.44\\
HRP dcca & $0.09 \pm 0.29$ & -0.02 & 20.67 & 0.04 & 0.062 & 2.57 & 0.29 & 0.49 & 0.54\\
HRP dpcca & $0.08 \pm 0.28$ & -0.03 & 22.0 & 0.03 & 0.055 & 2.72 & 0.31 & 0.46 & 0.52\\
NetMod cov & $0.09 \pm 0.22$ & 0 & 28.72 & 0.01 & 0.045 & 3.59 & 0.4 & 0.36 & 0.41\\
NetMod dcca & $0.09 \pm 0.26$ & 0.04 & 26.38 & 0.01 & 0.066 & 3.08 & 0.35 & 0.43 & 0.48\\
NetMod dpcca & $0.09 \pm 0.26$ & 0.05 & 27.32 & 0.01 & 0.064 & 3.11 & 0.36 & 0.41 & 0.46\\
CLA cov & $0.09 \pm 0.51$ & 0.1 & -11.12 & 0.31 & 0.187 & 1.86 & 0.18 & 0.82 & 0.92\\
\hline
\end{tabular}
\end{center}\label{tab:garch}\end{table*}

\begin{figure*}
\centering
  \includegraphics[width=\linewidth]{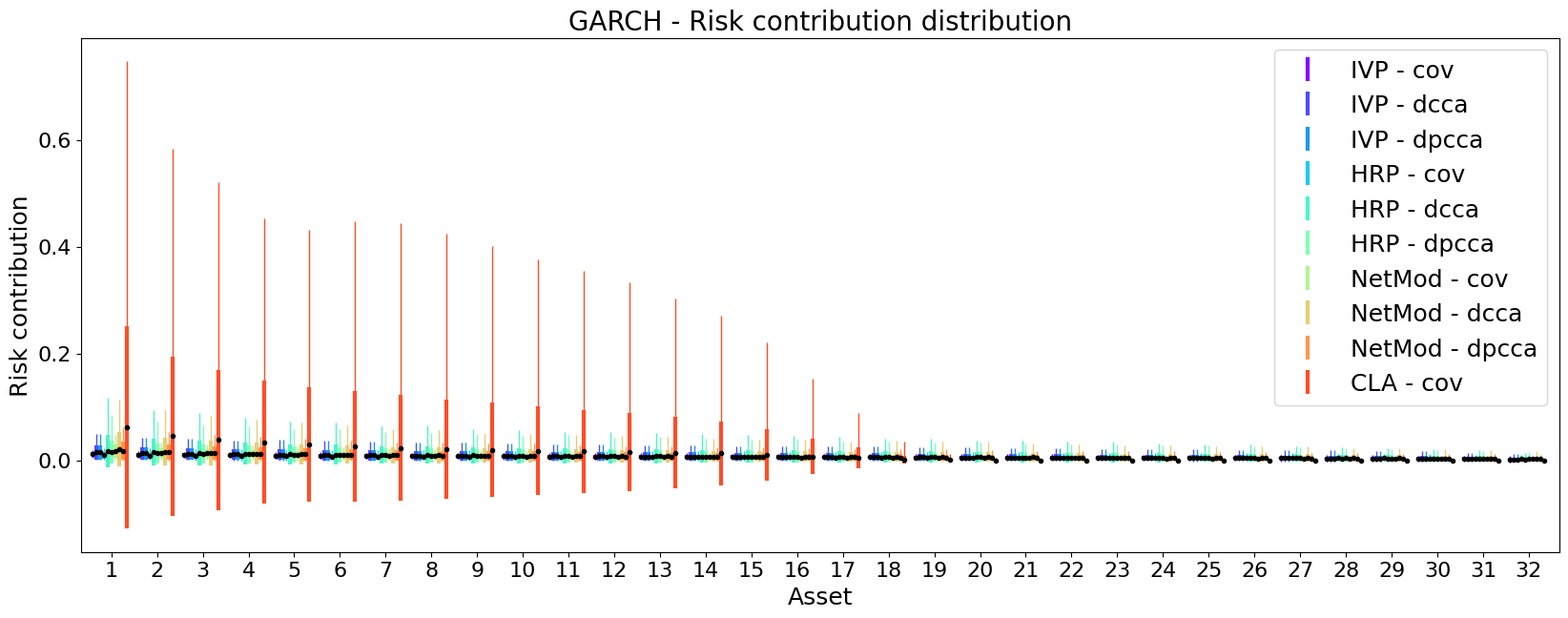}
\caption{GARCH: Risk contribution distribution}
\label{fig:garch-RC}
\end{figure*}

\begin{figure*}
\centering
  \includegraphics[width=\linewidth]{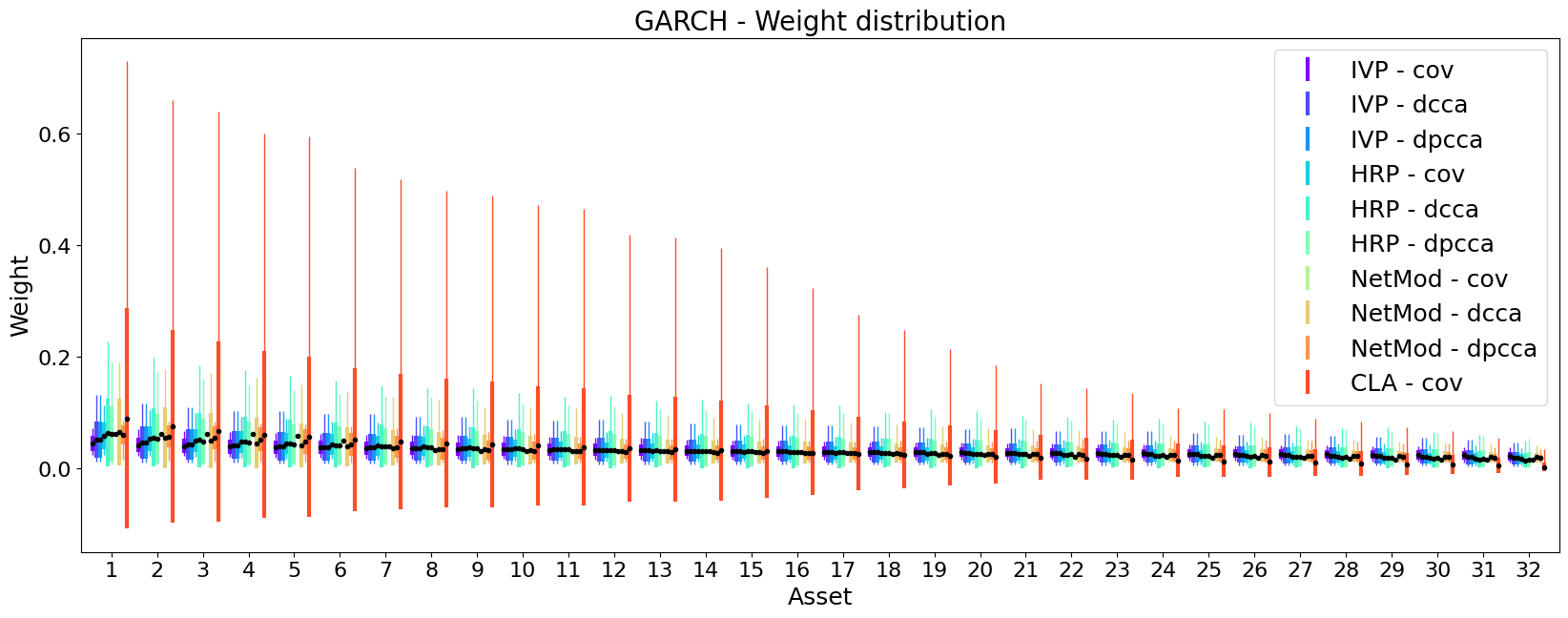}
\caption{GARCH: Weight distribution}
\label{fig:garch-weight-dist}
\end{figure*}

\begin{figure*}
\centering
  \includegraphics[width=.5\linewidth]{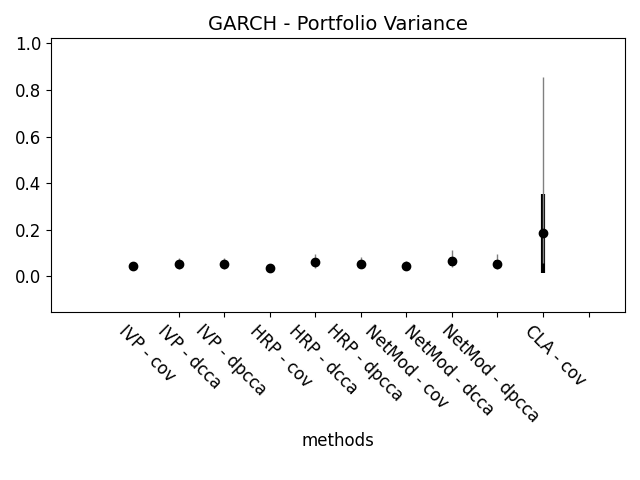}
\caption{GARCH: portfolio variance}
\label{fig:garch_pf}
\end{figure*}

Table \ref{tab:garch} shows the performances of different allocation schemes obtained running the GARCH simulations. NetMod schemes outperform others, since they offer higher return improvements, lower NHHIs, higher DRs, higher SRs, lower VaR and CVaR values. Other metrics provide comparable results. Again, worst results are those obtained for CLA. This is confirmed by looking at Figures \ref{fig:garch-RC}--\ref{fig:garch_pf}, showing, respectively the risk contribution distributions, weight distributions and portfolio variances.

\subsection{ARFIMA simulations}

\begin{table*}
\caption{ARFIMA simulation - results}
\begin{center}
\footnotesize
\begin{tabular}{ | l | c | c | c | c | c | c | c | c | c |}
\hline
\textbf{method} &\textbf{daily ret} &\textbf{Impr} & \textbf{CLR} &  \textbf{NHHI} &\textbf{PV} & \textbf{DR} &  \textbf{SR}  &  \textbf{VaR}  &  \textbf{CVaR}\\
IVP cov & $0.01 \pm 0.14$ & 0 & -1.09 & 0 & 0.017 & 2.33 & 0.05 & 0.3 & 0.34\\
IVP dcca & $0.01 \pm 0.14$ & -0.34 & -2.12 & 0.01 & 0.02 & 2.16 & 0.04 & 0.32 & 0.36\\
IVP dpcca & $0.01 \pm 0.14$ & -0.34 & -2.12 & 0.01 & 0.02 & 2.16 & 0.04 & 0.32 & 0.36\\
HRP cov & $0.01 \pm 0.13$ & 0.3 & -1.13 & 0.01 & 0.016 & 2.42 & 0.05 & 0.29 & 0.34\\
HRP dcca & $0 \pm 0.16$ & 0.59 & -4.24 & 0.04 & 0.025 & 1.93 & 0.03 & 0.36 & 0.41\\
HRP dpcca & $0.01 \pm 0.15$ & -0.62 & -3.28 & 0.03 & 0.022 & 2.05 & 0.04 & 0.34 & 0.39\\
NetMod cov & $0.01 \pm 0.13$ & 0.59 & -0.55 & 0.01 & 0.017 & 2.45 & 0.06 & 0.29 & 0.33\\
NetMod dcca & $0.01 \pm 0.16$ & 1.23 & -0.98 & 0.01 & 0.025 & 2.2 & 0.07 & 0.34 & 0.38\\
NetMod dpcca & $0.01 \pm 0.16$ & 0.44 & -0.95 & 0.01 & 0.024 & 2.19 & 0.07 & 0.34 & 0.38\\
CLA cov & $0.01 \pm 0.31$ & -0.3 & -20.47 & 0.28 & 0.079 & 1.36 & 0.02 & 0.69 & 0.83\\
\hline
\end{tabular}
\end{center}\label{tab:arfima}\end{table*}

\begin{figure*}
\centering
  \includegraphics[width=\linewidth]{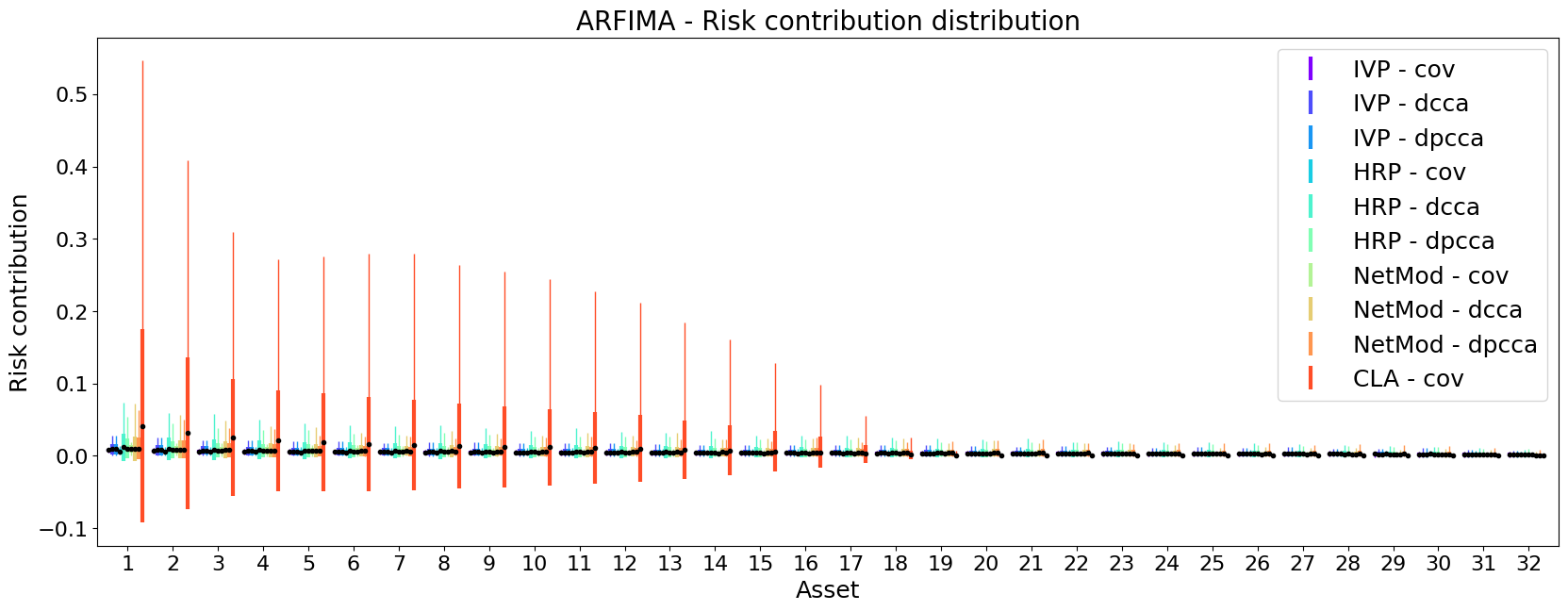}
\caption{ARFIMA: Risk contribution distribution}
\label{fig:arfima-RC}
\end{figure*}

\begin{figure*}
\centering
  \includegraphics[width=\linewidth]{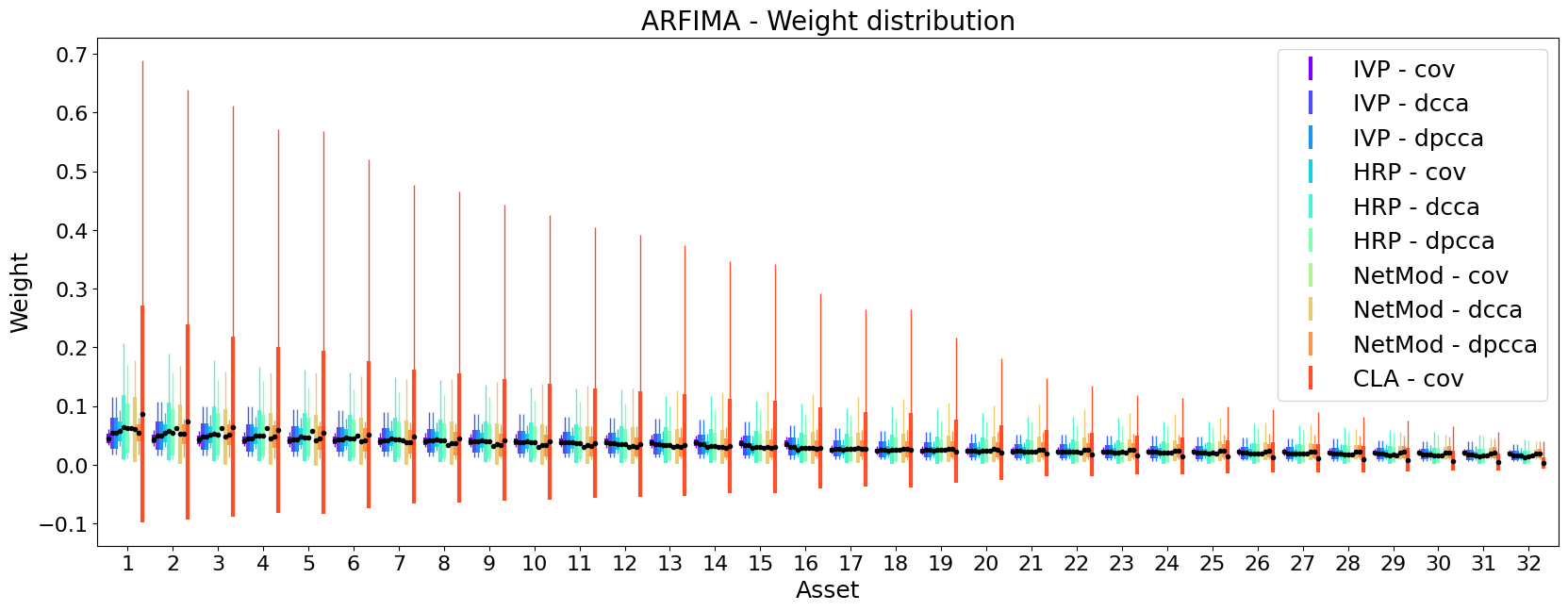}
\caption{ARFIMA: Weight distribution}
\label{fig:arfima-weight-dist}
\end{figure*}

\begin{figure*}
\centering
  \includegraphics[width=.5\linewidth]{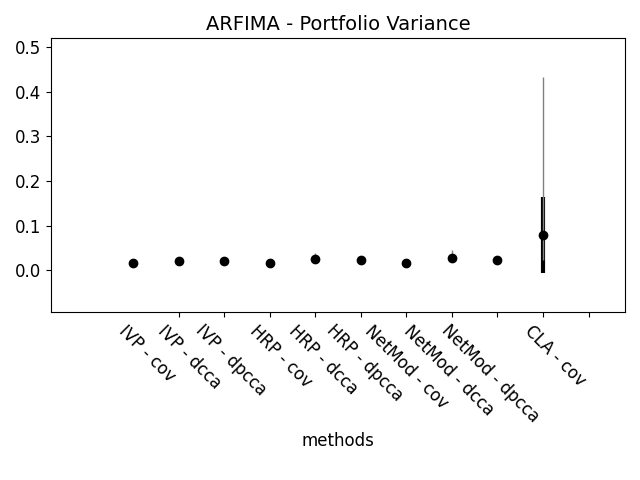}
\caption{ARFIMA: porfolio variance}
\label{fig:arfima-pf}
\end{figure*}

Table \ref{tab:arfima} shows the metrics obtained during the ARFIMA simulations. In this case, we have similar results for all the employed methods, except for CLA, that performs worse than other approaches, i.e., higher NHHI, higher average PV, lower DR, lower SR, higher VaR and CVaR. IVP and NetMod show better performance if we look at the NHHI. Positive daily return improvements are obtained for NetMod, especially ``NetMod dpcca". 
NetMod provides higher SR values as well, especially the ``dcca" and ``dpcca" variants. This happens without an increment on the standard deviations of the returns or increments of VaR, CVaR, which are instead lower that those of HRP methods.

Figures \ref{fig:arfima-RC}-\ref{fig:arfima-weight-dist} show the distributions of risk contributions and weights allocations. CLA assigns higher risks and weights to some specific assets, with a wide standard deviation. Instead, other approaches distribute risks and weights among multiple (all) assets. In general, a higher distribution is obtained when the ``dcca" variants are used. Such approaches also show wider standard deviations w.r.t.~``cov" variants, as well as maximum and minimum values. 
Moreover, HRP methods have higher variances with respect to NetMod and IVP.
In any case, the measured PV is limited for all methods (except for CLA), as shown in Figure \ref{fig:arfima-pf}.

\subsection{ARFIMA and correlation with shocks simulations}

\begin{table*}
\caption{ARFIMA and correlation with shocks simulation - results}
\begin{center}
\footnotesize
\begin{tabular}{ | l | c | c | c | c | c | c | c | c | c |}
\hline
\textbf{method} &\textbf{daily ret} &\textbf{Impr} & \textbf{CLR} &  \textbf{NHHI} &\textbf{PV} & \textbf{DR} &  \textbf{SR}  &  \textbf{VaR}  &  \textbf{CVaR}\\
IVP cov & $0.01 \pm 0.14$ & 0 & 1.03 & 0 & 0.017 & 2.32 & 0.09 & 0.29 & 0.33\\
IVP dcca & $0.01 \pm 0.15$ & -0.28 & 0.23 & 0.01 & 0.02 & 2.14 & 0.08 & 0.31 & 0.35\\
IVP dpcca & $0.01 \pm 0.15$ & -0.28 & 0.23 & 0.01 & 0.02 & 2.14 & 0.08 & 0.31 & 0.35\\
HRP cov & $0.01 \pm 0.13$ & -0.22 & 1.21 & 0.01 & 0.016 & 2.41 & 0.09 & 0.28 & 0.32\\
HRP dcca & $0.01 \pm 0.17$ & -0.2 & -1.08 & 0.04 & 0.025 & 1.92 & 0.07 & 0.36 & 0.4\\
HRP dpcca & $0.01 \pm 0.16$ & 0.4 & -0.64 & 0.03 & 0.022 & 2.04 & 0.07 & 0.33 & 0.37\\
NetMod cov & $0.01 \pm 0.13$ & 0.5 & 1.83 & 0.01 & 0.017 & 2.45 & 0.1 & 0.29 & 0.32\\
NetMod dcca & $0.01 \pm 0.16$ & 1.8 & 0.92 & 0.01 & 0.025 & 2.19 & 0.1 & 0.34 & 0.38\\
NetMod dpcca & $0.01 \pm 0.16$ & 0.62 & 0.78 & 0.01 & 0.024 & 2.19 & 0.09 & 0.33 & 0.37\\
CLA cov & $0.02 \pm 0.32$ & 5.49 & -17.15 & 0.28 & 0.078 & 1.35 & 0.06 & 0.68 & 0.82\\
\hline
\end{tabular}
\end{center}\label{tab:arfimashocksv2}\end{table*}

\begin{figure*}
\centering
  \includegraphics[width=\linewidth]{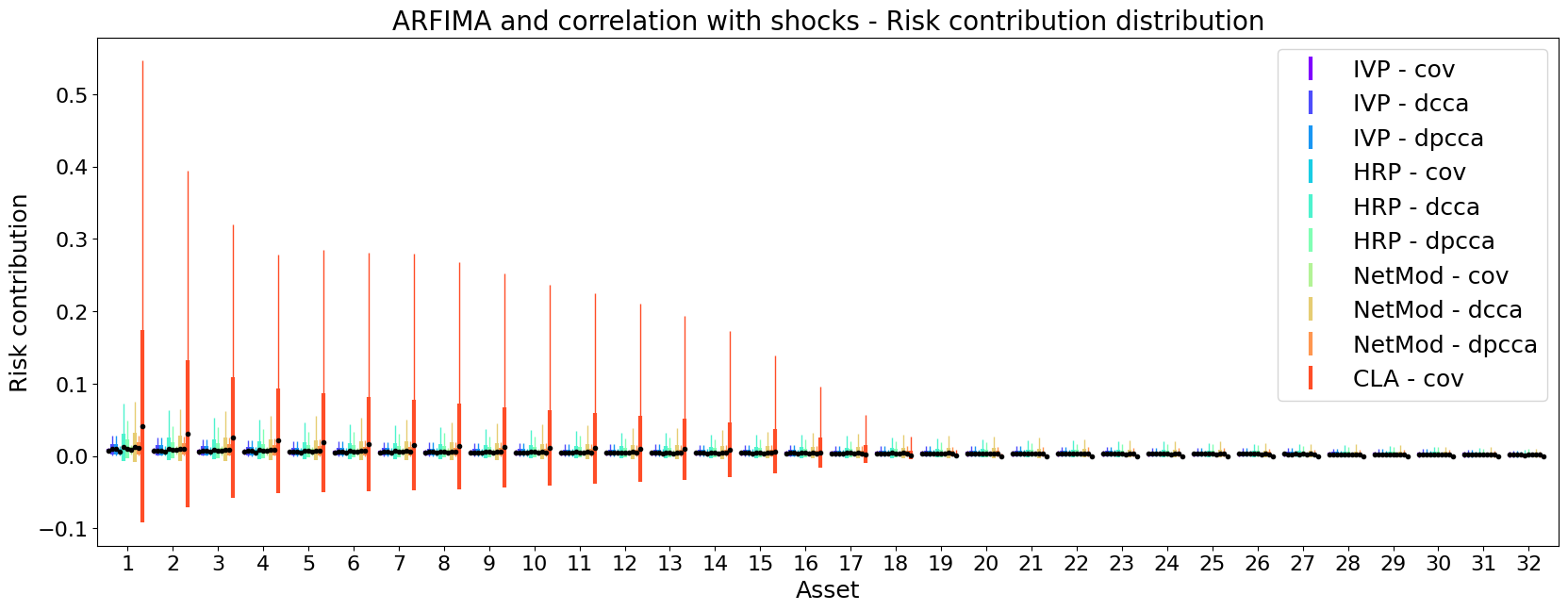}
\caption{ARFIMA and correlation with shocks: Risk contribution distribution}
\label{fig:arfimaShockv2-RC}
\end{figure*}

\begin{figure*}
\centering
  \includegraphics[width=\linewidth]{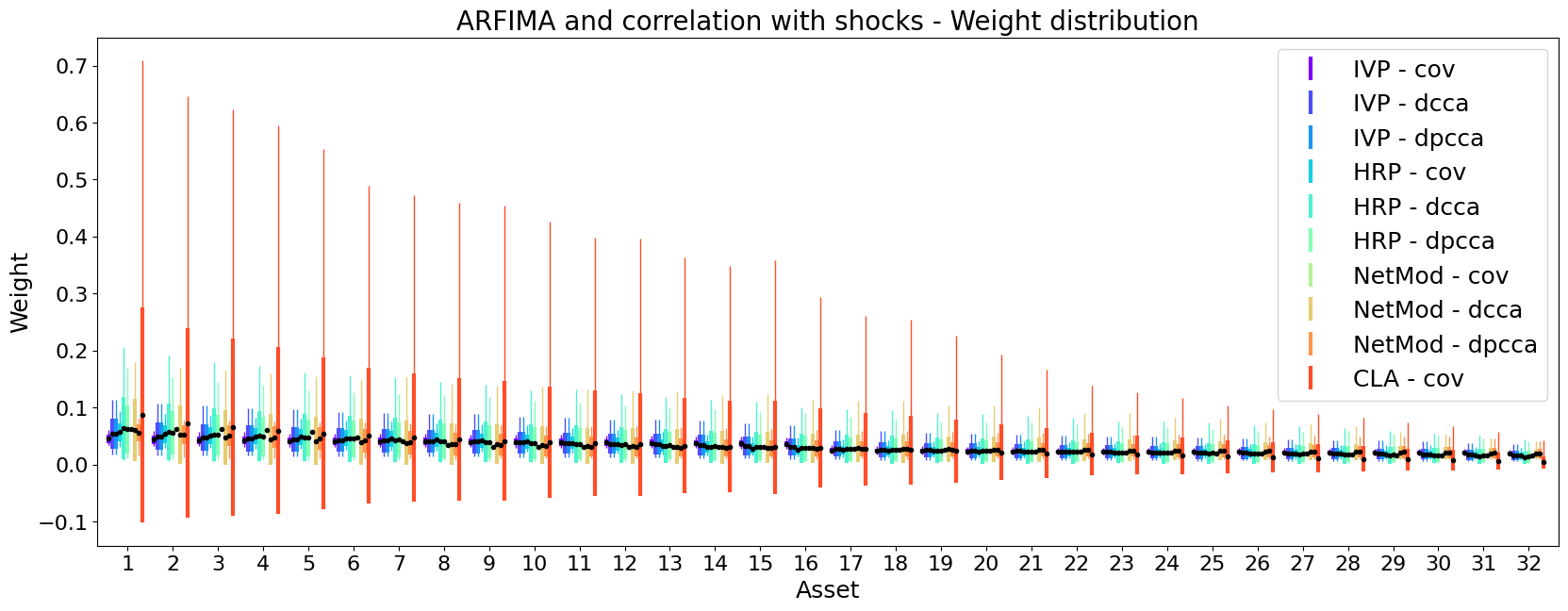}
\caption{ARFIMA and correlation with shocks: Weight distribution}
\label{fig:arfimaShockv2-weight-dist}
\end{figure*}

\begin{figure*}
\centering
  \includegraphics[width=.5\linewidth]{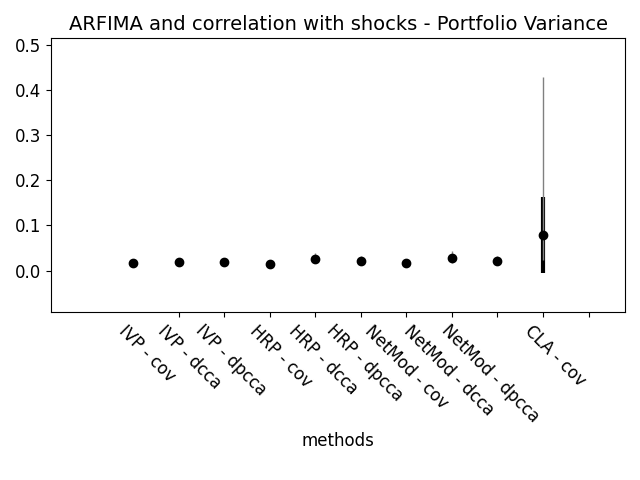}
\caption{ARFIMA and correlation with shocks: portfolio variance}
\label{fig:arfimaShockv2_pf}
\end{figure*}

Table \ref{tab:arfimashocksv2} shows results for the simulations where assets returns were generated through ARFIMA processes mixed with some temporal adjustments to further increase correlation, and where some random shocks were introduced. As in other contexts, CLA is the worst method. Among other approaches, it seems that better results are obtained when ``cov'' schemes are employed in this case. 
As concerns the returns, NetMod outperforms other approaches. 

A comparison between results related to this type of simulation (Table \ref{tab:arfimashocksv2}) and those obtained for ARFIMA (Table \ref{tab:arfima}), shows that in this type of simulation ``NetMod dcca" has higher improvements, with respect to others schemes. Moreover, while the CLR is always negative in ARFIMA, here the NetMod approaches have positive CLRs. NetMod approaches also increase the measured SR, while VaR and CVaR remain stable w.r.t.~ARFIMA.

Figures \ref{fig:arfimaShockv2-RC}-\ref{fig:arfimaShockv2_pf} show the risk contributions, weight allocations and portfolio variance, as for other types simulations. The general trends are confirmed also in this case. We can also appreciate a slight increment in the variability of risk contribution and weight allocations for the ``dcca" schemes.

\section{Backtest Results}\label{sec:backtest}

\begin{table*}
\caption{Backtest simulation - results}
\begin{center}
\footnotesize
\begin{tabular}{ | l | c | c | c | c | c | c | c | c | c |}
\hline
\textbf{method} &\textbf{daily ret} &\textbf{Impr} & \textbf{CLR} &  \textbf{NHHI} &\textbf{PV} & \textbf{DR} &  \textbf{SR}  &  \textbf{VaR}  &  \textbf{CVaR}\\
IVP cov & $0 \pm 0.01$ & 0 & 0.35 & 0.06 & 0 & 0.03 & 0.06 & 0.04 & 0.07\\
IVP dcca & $0 \pm 0.01$ & -0.05 & 0.33 & 0.07 & 0 & 0.03 & 0.06 & 0.05 & 0.07\\
IVP dpcca & $0 \pm 0.01$ & -0.05 & 0.33 & 0.07 & 0 & 0.03 & 0.06 & 0.05 & 0.07\\
HRP cov & $0 \pm 0.01$ & 0.11 & 0.41 & 0.14 & 0 & 0.03 & 0.09 & 0.03 & 0.05\\
HRP dcca & $0 \pm 0.01$ & 0.05 & 0.39 & 0.13 & 0 & 0.03 & 0.08 & 0.04 & 0.05\\
HRP dpcca & $0 \pm 0.01$ & -0.17 & 0.28 & 0.1 & 0 & 0.03 & 0.05 & 0.05 & 0.07\\
NetMod cov & $0 \pm 0.01$ & 0.45 & 0.53 & 0.06 & 0 & 0.04 & 0.09 & 0.04 & 0.06\\
NetMod dcca & $0 \pm 0.02$ & 0.23 & 0.41 & 0.02 & 0 & 0.04 & 0.06 & 0.05 & 0.08\\
NetMod dpcca & $0 \pm 0.02$ & 0.43 & 0.49 & 0.02 & 0 & 0.04 & 0.07 & 0.06 & 0.09\\
CLA cov & $0 \pm 0.02$ & 0.06 & 0.35 & 0.93 & 0 & 0.02 & 0.05 & 0.06 & 0.08\\
\hline
\end{tabular}
\end{center}\label{tab:backtest}\end{table*}

\begin{figure*}
\centering
  \includegraphics[width=\linewidth]{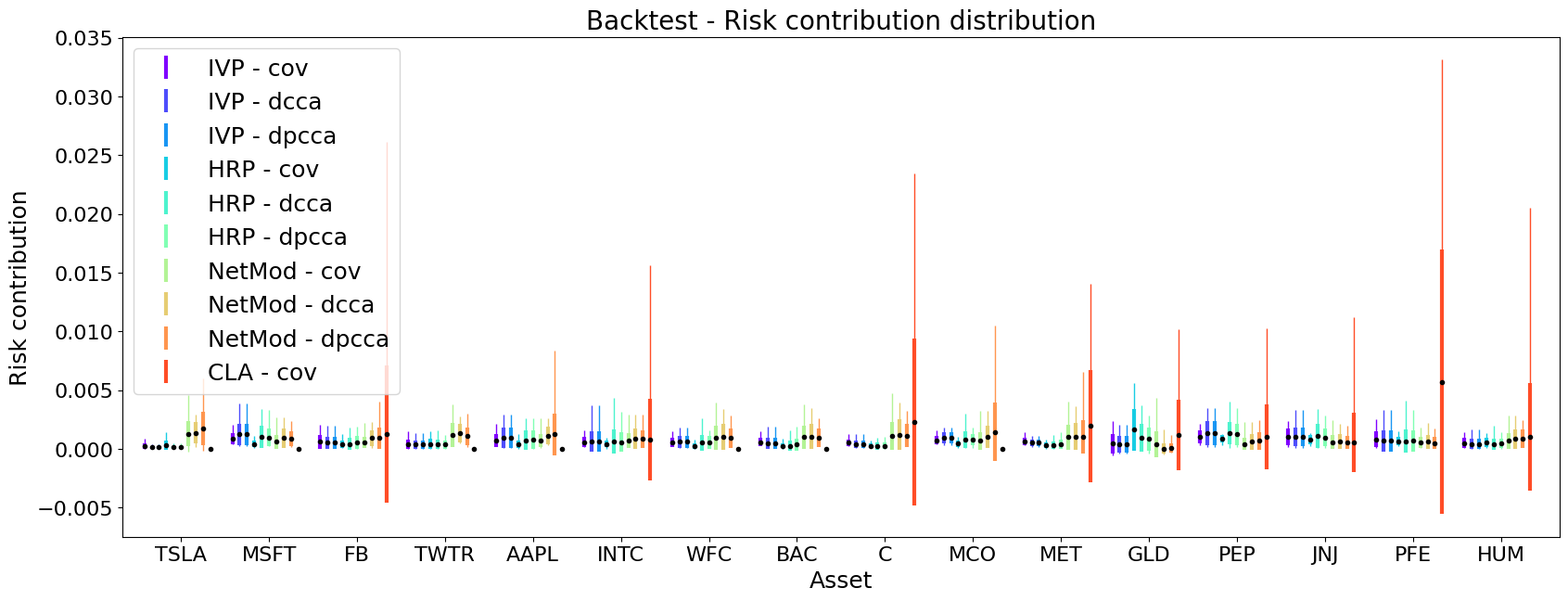}
\caption{Backtest: Risk contribution distribution}
\label{fig:backtest-RC}
\end{figure*}

\begin{figure*}
\centering
  \includegraphics[width=\linewidth]{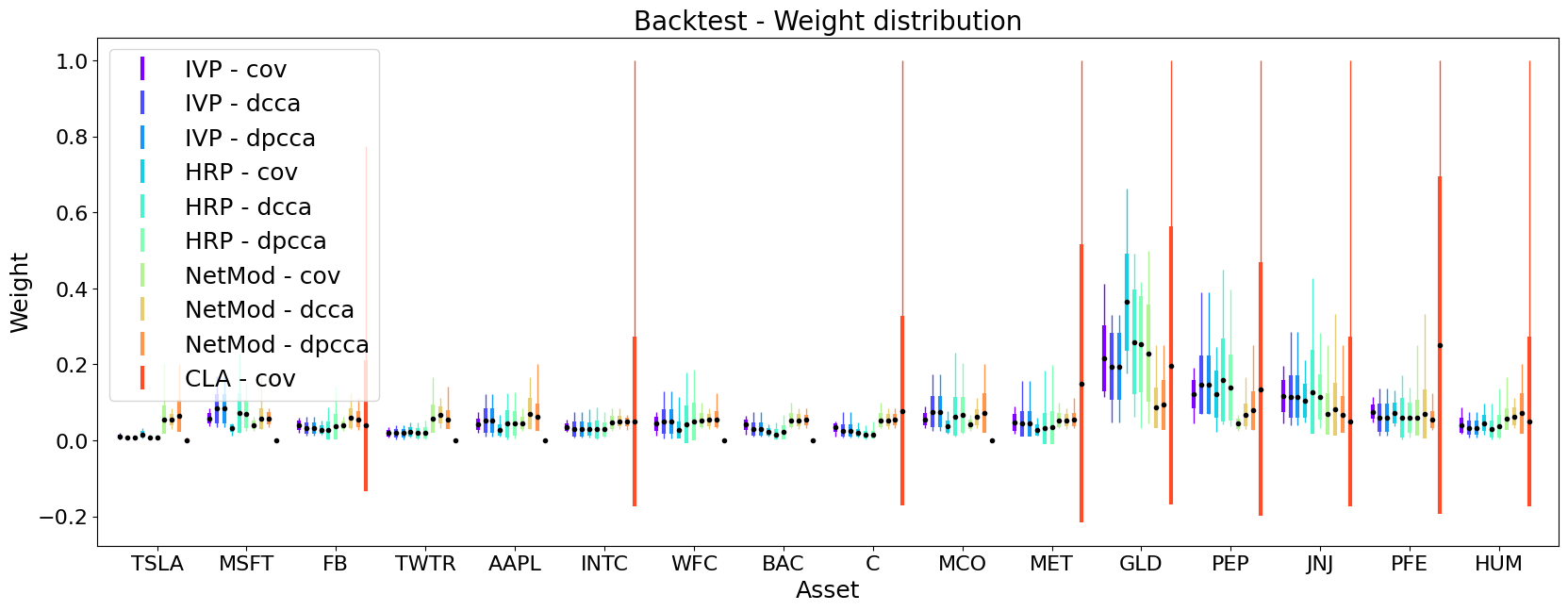}
\caption{Backtest: Weight distribution}
\label{fig:backtest-weight-dist}
\end{figure*}

\begin{figure*}
    \centering
    \includegraphics[width=.5\linewidth]{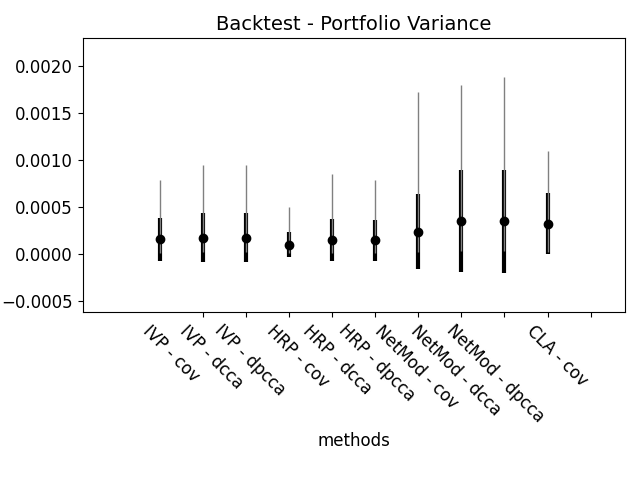}
    \caption{Backtest: portfolio variance}
    \label{fig:backtest-pf}
\end{figure*}

The outcomes obtained with the previous simulation studies are confirmed when backtests are performed. 
As shown in Table \ref{tab:backtest}, NetMod schemes perform better than others, since also in this case they show higher daily return improvements, higher CLRs, lower (better) NHHIs, higher (better) DRs, higher (better) SRs. As to VaR and CVar, similar results are obtained for all the schemes, with NetMod slightly worse than others. In these cases, all the values of PV are smaller than $10^{-4}$. 
By looking at the improvements on the returns, CLRs and other metrics, it seems that the ``dpcca" variants are the best ones.

Risk contributions and weight allocations are reported in Figures \ref{fig:backtest-RC}-\ref{fig:backtest-weight-dist}.
It is interesting to observe that all approaches tend to assign a high weight to GLD, which is a stable asset. 
In particular, IVP and HRP assign to GLD a weight significantly higher than other weights. This results in a corresponding higher risk contribution, which can be appreciated especially in ``HRP cov". CLA assigns positive weights to a limited amount of assets. As concerns NetMod schemes, 
in many cases, the community detection algorithm often puts GLD as a singleton asset, different than others.
Thus, ``NetMod cov" assigns high weight to GLD w.r.t.~others, which are nonetheless not null and higher than in IVP and HRP schemes. 
In ``NetMod cov", this results in a limited risk contribution associated with GLD. 
This aspect is even more evident in ``NetMod dcca", where the weight associated with other assets is higher. 
As a consequence, the risk contribution of GLD in ``NetMod dcca" is minimal.
This happens since the ``NetMod dcca" variant tends to create smaller communities than the ``NetMod cov" approach. Thus, the weights are more distributed among different sets of assets. This outcome is confirmed by the NHHI measures reported in Table \ref{tab:backtest}, where ``NetMod dcca" (and ``NetMod dpcca", similarly) has a way lower value than other schemes, thus resulting in a more shared allocation and competitive market, with no dominant assets.

\section{Discussion}
\label{sec:disc}

\subsection{Summary of Obtained Results}
Table \ref{tab:summary} provides a summary of the results, discussed in the previous sections, to facilitate a general overview of the outcomes. In particular, each row of the table shows results related to specific metrics. Results about the returns, shown in the previous tables (i.e., daily ret, Impr) have been aggregated into a single one (Ret). Moreover, PV has been omitted here, since it did not show differences among the approaches in almost all the scenarios, except for CLA, as already discussed in detail. Each column of the table focuses on a specific type of simulation or backtest. For each metrics and each type of simulation (or backtest), the table reports those approaches that showed the best results. In general, a specific allocation scheme with the related measure of correlation variant is reported (e.g., NetMod cov) which means that the specific configuration provided the best results. In other cases, only the name of an allocation scheme is shown (e.g., NetMod), meaning that the particular scheme worked better than others regardless of the correlation measure. Similarly, when only the correlation metrics is reported, it means that the best results have been obtained with that specific metrics, regardless of the allocation scheme. Configurations are reported in brackets when some slight, not significant improvements have been noticed. Finally, void cells signify that no particular winner is available for that metrics in that scenario.

\begin{table*}
\caption{Summary of results: best approaches based on specific metrics and simulation typology}
\begin{center}
\footnotesize
\begin{tabular}{ | l | p{.11\linewidth} | p{.1\linewidth} | p{.11\linewidth} | p{.12\linewidth} | p{.12\linewidth} | p{.12\linewidth} | }
\hline
\textbf{metrics} & \textbf{Gaussian} & \textbf{GBM} & \textbf{GARCH} & \textbf{ARFIMA} & \textbf{ARFIMA w shocks}  & \textbf{Backtest} \\
\hline
\textbf{Ret} & NetMod cov & NetMod dpcca & NetMod dpcca & NetMod dcca & NetMod dcca &  NetMod (cov) \\
\hline
\textbf{CLR} & NetMod cov &     & NetMod cov & NetMod (dcca) & NetMod cov &  NetMod (cov) \\  
\hline
\textbf{NHHI} &  & NetMod (cov) &    & & & NetMod (dcca, dpcca) \\
\hline
\textbf{DR} & cov (HRP, NetMod) & NetMod (cov) & NetMod cov & NetMod cov & NetMod cov & NetMod \\
\hline
\textbf{SR}  & NetMod, HRP & NetMod (dpcca) & NetMod cov & NetMod (dcca, dpcca) & NetMod &  NetMod, HRP (cov) \\  
\hline
\textbf{VaR} & (cov) & (IVP, HRP) &  NetMod cov & IVP cov & cov & HRP cov \\ 
\hline
\textbf{CVaR} & (cov) & (IVP, HRP) & NetMod cov & cov (NetMod) & cov (HRP, NetMod) & HRP (cov, dcca) \\
\hline
\end{tabular}
\end{center}\label{tab:summary}\end{table*}

The decision on which allocation scheme performs better than others should be taken by looking at all the metrics, considered together. All of these metrics focus on a particular aspect of the allocation, and some of them, while extremely popular in the financial analysis sector, received critiques in the literature. For example, SR is one of the most common metrics used to evaluate the performance of investments. However, it has been recognized that SR values are accentuated by investments that do not have a normal distribution of returns, and in general, it is subject to estimation errors that can be substantial in some cases \cite{doi:10.2469/faj.v58.n4.2453}. Similarly, as a further example, VaR is by far another leading measure of portfolio allocation analysis and it is widely used in major banks and financial institutions. But also in this case, this metrics is not exempt from criticism. In particular, it has been recognized that this measure can discourage diversification \cite{RePEc:bla:mathfi:v:9:y:1999:i:3:p:203-228}.
Thus, a wider overview of the available metrics should give some insights on the performance of the considered schemes. 

The table clearly shows two important results. First, NetMod outperforms other portfolio allocation strategies in all the considered scenarios. In a few cases, other schemes provide better VaR (and CVaR in some cases), but differences are not so important.

Second, it seems that a naive replacement of the classic Pearson's correlation with DCCA or DPCCA does not provide significant benefits in many scenarios. The classic Pearson’s correlation performs better than DCCA and DPCCA in various situations, while the latter metrics might perform better when data traces are highly correlated and non stationary (e.g. ARFIMA). Thus, some further investigation might be needed in this sense.

\subsection{NetMod Communities Obtained when Varying Threshold and Correlation Metrics}

\begin{figure*}
    \centering
    \begin{minipage}{.3\textwidth}
    \includegraphics[width=\textwidth]{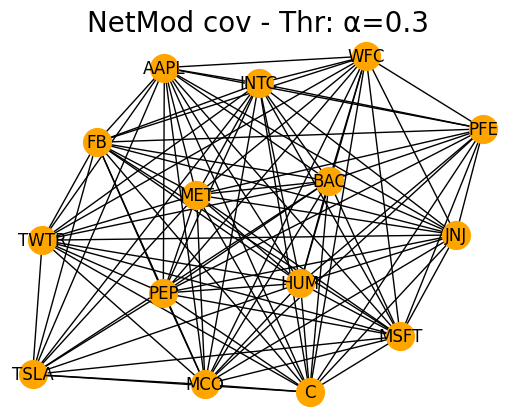}
    \end{minipage}
    \begin{minipage}{.3\textwidth}
    \includegraphics[width=\textwidth]{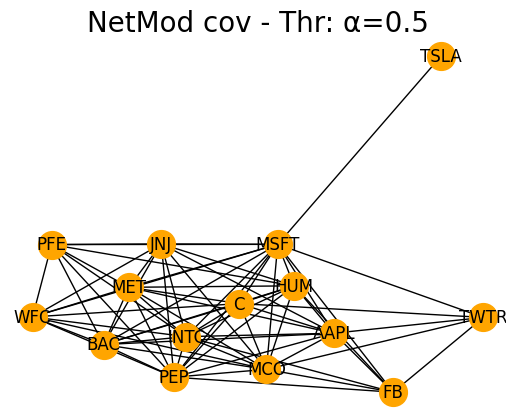}
    \end{minipage}
    \begin{minipage}{.3\textwidth}
    \includegraphics[width=\textwidth]{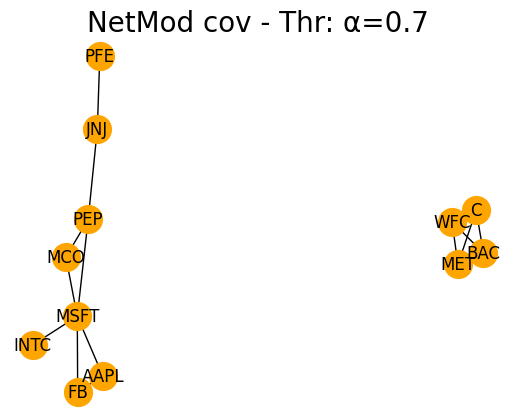}
    \end{minipage}
    \begin{minipage}{.3\textwidth}
    {\footnotesize
    c1: \{GLD\}\\
    c2: \{all others\}}
    \end{minipage}
    \begin{minipage}{.3\textwidth}
    {\footnotesize
    c1: \{INTC, WFC, BAC, \\
    \hspace*{5mm}MET, PEP, JNJ, PFE\}\\
    c2: \{TSLA, MSFT,\\ 
    \hspace*{5mm}FB, TWTR\}\\
    c3:  \{GLD\}
    }
    \end{minipage}
    \begin{minipage}{.3\textwidth}
    c1: \{MSFT, FB, AAPL,\\ 
    \hspace*{5mm}INTC\}\\
    c2: \{MCO, PEP, JNJ, \\
    \hspace*{5mm}PFE\}\\
    c3: \{WFC, BAC, C, MET\}\\
    Other assets isolated
    \vspace{0.3cm}
    \end{minipage}
    \begin{minipage}{.3\textwidth}
    \includegraphics[width=\textwidth]{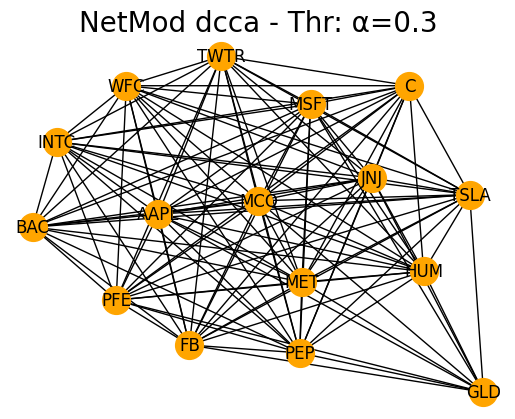}
    \end{minipage}
    \begin{minipage}{.3\textwidth}
    \includegraphics[width=\textwidth]{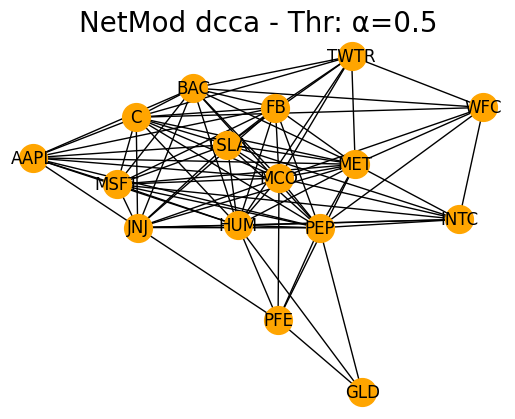}
    \end{minipage}
    \begin{minipage}{.3\textwidth}
    \includegraphics[width=\textwidth]{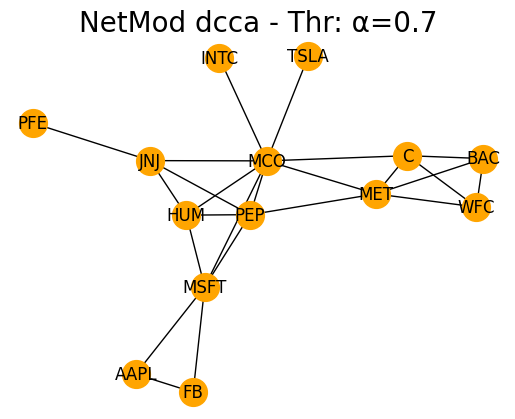}
    \end{minipage}
    \begin{minipage}{.3\textwidth}
    {\footnotesize
    c1: \{all assets\}}
    \end{minipage}
    \begin{minipage}{.3\textwidth}
    {\footnotesize
    c1: \{TSLA, MSFT, AAPL,\\
    \hspace*{5mm}INTC, JNJ\}\\
    c2: \{FB, TWTR, C, MCO,\\
    \hspace*{5mm}MET, WFC,BAC\}\\
    c3:  \{GLD, PEP, PFE, \\
    \hspace*{5mm}HUM\}
    }
    \end{minipage}
    \begin{minipage}{.3\textwidth}
    c1: \{TSLA, INTC, MCO\\ 
    \hspace*{5mm}PEP, PFE, HUM\}\\
    c2: \{MSFT, FB, AAPL\}\\
    c3: \{WFC, BAC, C, MET\}\\
    Other assets isolated
    \vspace{0.3cm}
    \end{minipage}
    \begin{minipage}{.3\textwidth}
    \includegraphics[width=\textwidth]{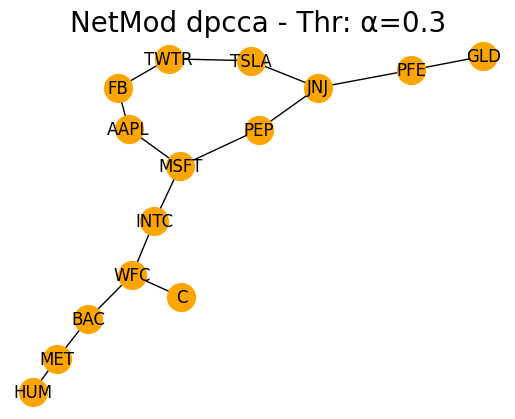}
    \end{minipage}
    \begin{minipage}{.3\textwidth}
    \includegraphics[width=\textwidth]{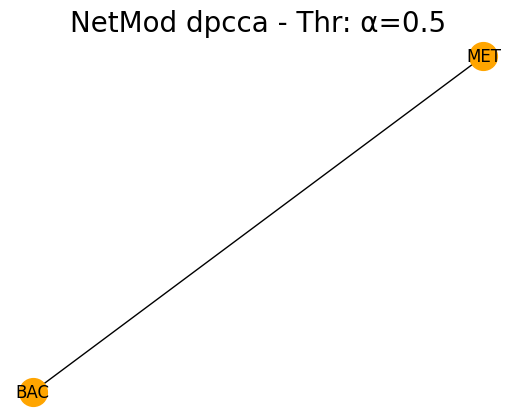}
    \end{minipage}
    \begin{minipage}{.3\textwidth}
    \includegraphics[width=\textwidth]{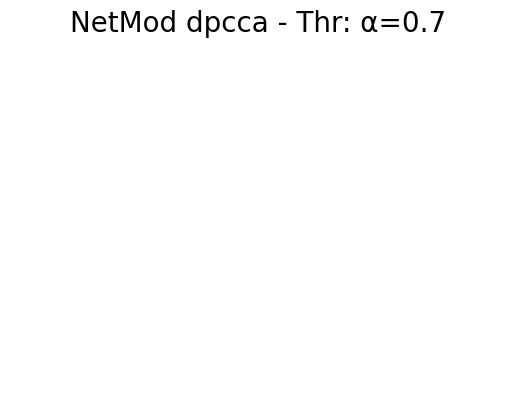}
    \end{minipage}
    \begin{minipage}{.3\textwidth}
    {\footnotesize
    c1: \{MSFT, PEP\}\\
    c2: \{FB, TWTR, AAPL\}\\
    c3: \{TSLA, GLD, JNJ, \\
    \hspace*{5mm}PFE\}\\
    c4: \{INTC, WFC, C\}\\
    c5: \{BAC, MET, HUM\}\\
    c6: \{MCO\}
    }
    \end{minipage}
    \begin{minipage}{.3\textwidth}
    {\footnotesize
    c1: \{BAC, MET\}\\
    Other assets isolated
    }
    \end{minipage}
    \begin{minipage}{.3\textwidth}
    All assets isolated
    \vspace{0.3cm}
    \end{minipage}
    \caption{MetMod in backtest: networks formed by links whose edges have weight above the threshold and obtained communities, based on the used correlation metrics}
    \label{fig:communities}
\end{figure*}

Figure \ref{fig:communities} shows the networks generated by the NetMod scheme in the backtest, in a specific window of $64$ observations. In particular, each reported network depends on the threshold $\alpha$ that is used to create links among nodes in the network (as discussed in Section \ref{sec:netmod}) and on the specific correlation metrics in use. Below each network, it is reported the list of communities of assets obtained using the Louvain algorithm to measure the network modularity. 
Each row shows three different networks when varying $\alpha$, while keeping the same correlation metrics. Each column shows the networks obtained using the different correlation metrics while keeping fixed the value of $\alpha$.

The figure clarifies two main aspects: i) as expected, the higher the threshold, the less connected the network; ii) the threshold $\alpha$ has a different impact on the generation of the networks depending on the used correlation metrics. DPCCA is more threshold sensitive. This suggests that, instead of using a fixed value for $\alpha$, equal for all the correlation variants as done in this evaluation study, the scheme should be accompanied by preliminary test and validation phases to properly tune the hyper-parameter $\alpha$. This investigation is left as a future work.

\section{Conclusions}
\label{sec:conc}

In this work, it has been shown that modeling, simulation and data science techniques can be extremely useful to evaluate complex scenarios and to design novel approaches in financial applications, such as portfolio analysis and allocation. 
In fact, first, a detailed and variegated simulation framework allows performing what-if analyses and studying if the devised schemes perform well in different situations. 

Second, we have shown the feasibility of using alternative measures to describe and characterize data series. 
Here, we used the notions of DCCA and DPCCA as an alternative to the classic Pearson's correlation measure. In the portfolio allocation approaches, we naively replaced the original correlation with these metrics. While the obtained results do not outperform the classic correlation in many considered scenarios, we showed that a proper tuning of the hyper-parameters employed in the allocation schemes can have a strong influence on the final performance. This tuning should be accomplished through specific data analysis, once a specific correlation metrics is used.
The benefits of DCCA and DPCCA on portfolio allocation thus remain an open question. This aspect will be future work.
Another point worthy of mention is that a drawback of the Pearson measure is its inability to capture the directionality of the relationships between asset classes. Looking into directionality might be helpful for a better analysis of assets' interdependence and might provide interesting insights. Further research can be devoted to the study of measures to capture directionality and see if they can have some impact on asset community identification.

Third, we described a novel scheme, here called Naive Network Modularity based allocation (NetMod), that takes all the assets and, based on the employed correlation metrics, builds a complex network. 
Communities of assets are identified thanks to modularity. 
The approach then distributes investments among such communities. 
While simple, the approach outperforms the state of the art approaches in many situations. Possible improvements can go in the direction of not equally distributing the weights, but consider, for example, the variance or risk of the communities, with intra and inter-communities optimizations. 
Further investigation might be on the analysis of alternatives to the exploited Louvain method to detect communities. Examples worth of study are the Walktrap algorithm, Infomap, the Fast-greedy, and the Leading Eigenvalue \cite{commdetect}.

All this confirms the benefits of using data science and complex networks theories in portfolio management and financial applications. The obtained results foster the claim that there is room for improvement and thus, further research is needed. On the other hand, such applications can in turn foster novel research in the data and information science theoretical domains, trying to cope with the need to measure financial metrics and to design effective modeling and simulation techniques.

\section*{Declarations}
S. Ferretti declares that he has no conflict of interest.


\end{document}